\documentclass[12pt]{article}
\usepackage{a4wide,amssymb,cite}
\pdfoutput=1

\usepackage{a4wide,amssymb,graphicx}
\usepackage{epsfig}
\usepackage[usenames,dvipsnames]{color}
\usepackage{slashed}
\usepackage{amssymb,cite,graphicx}
\usepackage{slashed}
\usepackage{amsmath,bm,bbm}
\usepackage{amsfonts}
\usepackage[titletoc,title]{appendix}

\usepackage[small]{caption}
\usepackage[margin=1in]{geometry}
\usepackage[multiple]{footmisc}
\usepackage{mathtools}
\usepackage{slashed}
\usepackage[nottoc]{tocbibind}
\usepackage{xcolor}

\usepackage[compat=1.1.0]{tikz-feynman}

\newcommand{\be}{\begin{equation}}
\newcommand{\ee}{\end{equation}}
\newcommand{\bea}{\begin{eqnarray}}
\newcommand{\eea}{\end{eqnarray}}

\def\circa#1{\,\raise.3ex\hbox{$#1$\kern-.75em\lower1ex\hbox{$\sim$}}\,}

\begin{document}

\begin{titlepage}
%
%

%\rightline{CERN-PH-TH/2018-xxx}

%

\begin{centering}
\vspace{1cm}
{\Large {\bf Models for Self-resonant Dark Matter}} \\

\vspace{1.5cm}

{\bf Seong-Sik Kim$^{1,*}$, Hyun Min Lee$^{1,\dagger}$, and Bin Zhu$^{2,\ddagger}$}
%\\
\vspace{.5cm}

{\it $^1$Department of Physics, Chung-Ang University, Seoul 06974, Korea.} 
\vspace{.1cm} \\
{\it $^2$School of Physics, Yantai University, Yantai 264005, China.}
\vspace{0.5cm}\\
%\today

\end{centering}
\vspace{2cm}

\begin{abstract}
\noindent
We consider a new mechanism for enhancing the self-scattering and annihilation cross sections for dark matter with multiple components but without a light mediator. The lighter dark matter component plays a role of the $u$-channel pole
in the elastic co-scattering for dark matter, leading to a large self-scattering cross section and a Sommerfeld enhancement for semi-annihilation processes. Taking the effective theory approach for self-resonant dark matter, we present various combinations of multiple dark matter components with spins and parities, showing a $u$-channel pole in the co-scattering processes. Adopting dark photon and dark Higgs portals for self-resonant dark matter, we impose the relic density condition as well as indirect detection bounds on semi-annihilation channels with a Sommerfeld enhancement and  discuss potential signals for direct detection experiments. 

\end{abstract}

\vspace{5cm}

\begin{flushleft} 
$^*$Email: sskim.working@gmail.com \\   
$^\dagger$Email: hminlee@cau.ac.kr  \\
$^\ddagger$Email:  zhubin@mail.nankai.edu.cn 
\end{flushleft}

\end{titlepage}

\tableofcontents
\vspace{35pt}
\hrule

\section{Introduction}

%dark matter
Dark matter is a big mystery in astrophysics and particle physics, because the nature of particle dark matter or the composition of dark matter is completely unknown. However, we have been determining the dark matter abundance at cosmological scales with high precision from the Cosmic Microwave Background (CMB) and improving the information on the local density and velocity distribution of dark matter in galaxies. There have been a lot of efforts to test various scenarios for dark matter, such as Weakly Interacting Massive Particles (WIMPs), Axions, Strongly Interacting Massive Particles (SIMPs), Primordial Black Holes, etc. Therefore, it is plausible to see some hints for dark matter in a near future.

%SIDM, non-perturbative scattering
Non-perturbative effects for dark matter can be important to make a precise determination of the relic density for dark matter and indirect signals from dark matter annihilation \cite{hisano,cirelli,russell,AH,pospelov,cassel,lengo,slatyer,feng,blum,kai} and dark matter self-scattering \cite{smallscale1,smallscale2,yu,slatyer2,zhang,kai2,kang,felix}. In particular, in the presence of light mediators for dark matter, the Born approximation for dark matter scattering cross section breaks down due to  Coulomb-like singularities or $t$-channel resonances, so the inclusion of higher order processes  leads to the enhancement in the dark matter annihilation process, the so called Sommerfeld effects \cite{sommerfeld}. The non-perturbative effects with light mediators can render dark matter self-scattering enhanced and velocity-dependent to explain small-scale problems at galaxies \cite{sidm0,smallscale3,diversity,sidm}.

%model1
In this article, we present a comprehensive analysis of the new mechanism for enhancing the dark matter scattering cross section without light mediators, but with $u$-channel resonances, as proposed by the authors \cite{SRDM}. The new enhancement mechanism relies on the assumption that dark matter is composed of at least two components with comparable masses. Suppose that two scalar dark matter particles, $\phi_1$ and $\phi_2$, have a triple coupling such as $\phi_2|\phi_1|^2$ and satisfy an approximate mass relation, $m_2\lesssim 2m_1$. Then,  the lighter dark matter $\phi_1$ can play a role of the mediator with a zero effective mass for the $u$-channel elastic scattering between two dark matter particles \cite{SRDM}. This new possibility was dubbed ``self-resonant dark matter''.
We extend our previous analysis for the Sommerfeld factors for dark matter to the case with arbitrary partial wave processes and discuss the details on how the non-perturbative effects with $u$-channel resonances appear in the $2\to 2$ and $3\to 2$ annihilation processes for dark matter.

%model2
We also propose the effective theory for self-resonant dark matter by considering simple models for multi-component dark matter with different spins and parities. In this context, we show how the $u$-channel resonances appear in those general dark matter models. Furthermore, focusing on the model with two scalar dark matter particles and a dark photon mediator, we consider the thermal freeze-out for obtaining a correct total relic density for multi-component dark matter with equal or unequal fractions. We also discuss the implications of self-resonance dark matter for indirect detection such as from cosmic rays measurements and CMB data as well as for direct detection experiments.

%paper
The paper is organized as follows. 
We first present a simple model for two-component scalar dark matter and treat the $u$-channel resonances for dark matter scattering in a generalization of the Bethe-Salpeter formalism. Then, we use the non-perturbative enhancement for elastic co-scattering, $2\to 2$ and $3\to 2$ semi-annihilations for two-component dark matter. We continue to provide the effective theory for self-resonant dark matter with multi-components of various combinations of spins/parities and identify the cases where the $u$-channel resonances show up. Next, we discuss the phenomenology of self-resonant dark matter in the model for two-component scalar dark matter with a dark photon portal, such as the relic density for dark matter and the bounds from indirect and direct detection experiments. Finally, conclusions are drawn.

\section{New Sommerfeld effects}

We take a simple toy model for two-component dark matter with a triple coupling and discuss the $u$-channel resonance from a lighter component dark matter  appearing in the elastic co-scattering process. Developing the Bethe-Salpter formalism for the elastic scattering for dark matter with the $u$-channel resonance, we derive a Schr\"odinger-like equation for the BS wave function and show how to identify the Sommerfeld factor in this case.

\subsection{A model for two-component dark matter}

We first consider the Lagrangian for a complex scalar field $\phi_1$ and a real scalar field $\phi_2$, with a global or local $U(1)$ symmetry for $\phi_1$, in the following \cite{SRDM},
\bea
{\cal L}= |\partial_\mu\phi_1|^2 -m^2_1 |\phi_1|^2+\frac{1}{2}(\partial_\mu\phi_2)^2-\frac{1}{2} m^2_2 \phi^2_2 - 2g\,m_1 \phi_2 |\phi_1|^2. \label{scalardm}
\eea
For singlet scalar fields, we can add extra couplings such as $\phi^3_2$, $\phi_2^2|\phi_1|^2$, $|\phi_1|^4$ and $\phi^4_2$ but they are not relevant for our later discussion on the $u$-channel resonance.

In our model, there are elastic scattering processes between $\phi_1$ and $\phi_2$, such as $\phi_1\phi_2\rightarrow \phi_1 \phi_2$ or $\phi^*_1 \phi_2 \rightarrow \phi^*_1 \phi_2$, due to the triple coupling. 
In the absence of light mediators, henceforth we focus on the possibility of utilizing the $u$-channel resonances for  $\phi_1\phi_2\rightarrow \phi_1 \phi_2$ or $\phi^*_1 \phi_2 \rightarrow \phi^*_1 \phi_2$. We also comment on other processes such as $\phi_1\phi^{(*)}_1\rightarrow \phi_1 \phi^{(*)}_1$ or  $\phi_2\phi_2\rightarrow \phi_1 \phi^*_1$, due to the same triple coupling $g$.

\subsection{Elastic co-scattering}

We consider the elastic scattering, $\phi_1\phi_2\rightarrow \phi_1\phi_2$ and its complex conjugate, which can have a non-perturbative enhancement near $m_2\sim 2m_1$.

\begin{figure}[!t]
\begin{center}
\begin{tikzpicture} [baseline=($0.5*(ini1)+0.5*(ini2)$)]
	\begin{feynman} [inline=($0.5*(ini1)+0.5*(ini2)$), medium]
		\vertex (ini1) { \( \phi_1 (q) \) } 	;	\vertex [below = of ini1] (ini2) { \( \phi_2 (p) \) } ;
		\vertex [right = of ini1](int1) ;	\vertex [right = of ini2](int2) ;
		\vertex [right = of int1](f1) { \( \phi_1 (q') \) } ;	\vertex [right = of int2](f2) { \( \phi_2 (p') \) } ;
		
		\diagram*{
			(ini1) -- [charged scalar] (int1) -- [scalar] (f2),
			(ini2) -- [scalar] (int2) -- [charged scalar] (f1),
			(int1) -- [charged scalar, edge label = \( \phi_1 \)] (int2),
		};
	\end{feynman}
\end{tikzpicture}
\be
\begin{tikzpicture} [baseline=($0.5*(ini1)+0.5*(ini2)$)]
	\begin{feynman} [inline=($0.5*(ini1)+0.5*(ini2)$), medium]
		\vertex (ini1) { \( \phi_1 (q) \) } ;
		\vertex [below = of ini1] (ini2) { \( \phi_2 (p) \) } ;
		\vertex [right = of ini1](int1) ; \vertex [right = of ini2](int2) ;
		\vertex [right = 0.5cm of int1](f1) ; \vertex [right = 0.5cm of int2](f2) ;
		
		\diagram*{
			(ini1) -- [charged scalar] (int1) -- [scalar] (f1),
			(ini2) -- [scalar] (int2) -- [charged scalar] (f2),
			(int1) -- [charged scalar, edge label = \( \phi_1 \)] (int2),
		};
	\end{feynman}
	\end{tikzpicture}
	\cdots
	\begin{tikzpicture} [baseline=($0.5*(ini1)+0.5*(ini2)$)]
	\begin{feynman} [inline=($0.5*(ini1)+0.5*(ini2)$), medium]
		\vertex (ini1) ;
		\vertex [below = of ini1] (ini2) ;
		\vertex [right = 0.5cm of ini1] (u1) ; \vertex [right = of u1] (u2) ; \vertex [right = of u2] (u3) ;  \vertex [right = of u3] (f1) { \( \phi_1 (q') \) } ;
		\vertex [right = 0.5cm of ini2] (d1) ; \vertex [right = of d1] (d2) ; \vertex [right = of d2] (d3) ;  \vertex [right = of d3] (f2) { \( \phi_2 (p') \) } ;
				
		\diagram*{
			(ini1) -- [charged scalar, edge label = \( \phi_1 \)] (u1) -- [scalar, edge label = \( \phi_2 \)] (u2) -- [charged scalar, edge label = \( \phi_1 \)] (u3) -- [draw = none] (f1),
			(ini2) -- [scalar, edge label = \( \phi_2 \)] (d1) -- [charged scalar, edge label = \( \phi_1 \)] (d2) -- [scalar, edge label = \( \phi_2 \)] (d3) -- [draw = none] (f2),
			(u1) -- [charged scalar, edge label = \( \phi_1 \)] (d1), (u2) -- [anti charged scalar, edge label = \( \phi_1 \)] (d2), (u3) -- [charged scalar, edge label = \( \phi_1 \)] (d3), 
			(u3) --[scalar] (f2), (d3) -- [charged scalar] (f1),
		};
	\end{feynman}
	\end{tikzpicture}
\nonumber
\ee
\end{center}
\caption{Feynman diagrams ($u$-channel)  for $\phi_1\phi_2\to \phi_1\phi_2$  at tree level (above), and the corresponding ladder diagrams (below). }
\label{uFeyn}
\end{figure}
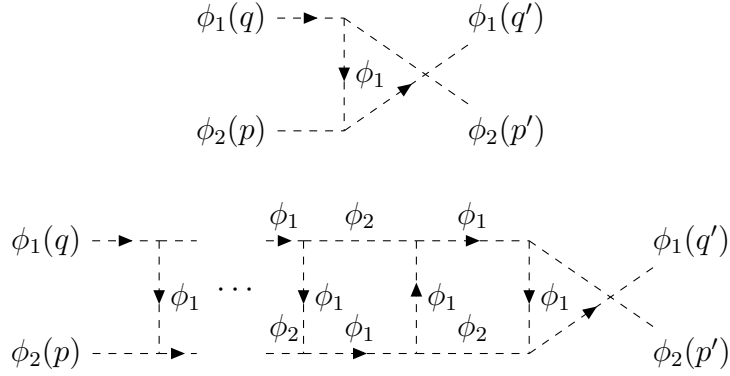

For the elastic scattering process, $\phi_1(q)\phi_2(p)\rightarrow \phi_1(q')\phi_2(p')$, 
the tree-level scattering amplitude ${\widetilde \Gamma}(p,q;p,q')$, as shown in the upper diagram in Fig.~\ref{uFeyn}, is given by
\bea
 {\widetilde \Gamma}(p,q;p',q') =-\frac{4g^2 m^2_1}{(p-q')^2-m^2_1}=\frac{4g^2 m^2_1}{|{\vec p}-{\vec q}'|^2+m^2_1-\omega^2}
\eea
with $\omega=p_0-q'_0$ being the energy exchange. 
Then, in the non-relativistic limit for initial particles, we get 
\bea
m^2_1-\omega^2&=&m^2_1-\Big(\sqrt{m^2_2+{\vec p}^2}-\sqrt{m^2_1+{\vec q}^{\prime 2}}\Big)^2 \nonumber \\
&\approx & m_2(2m_1-m_2)+\Big(-1+\frac{m_1}{m_2}\Big){\vec p}^2+\Big(-1+\frac{m_2}{m_1} \Big){\vec q}^{\prime 2}. \label{energytransfer}
\eea
We note that the energy exchange $\omega$ does not vanish in the non-relativistic limit, unlike the case for the elastic scattering between the same particles, so the process is not instantaneous.

As a result, the tree-level scattering amplitude ${\widetilde \Gamma}(p,q;p',q')$ becomes
\bea
 {\widetilde \Gamma}(p,q;p',q') \approx \frac{4g^2 m^2_1}{\Big(\sqrt{\frac{m_1}{m_2}} {\vec p}-\sqrt{\frac{m_2}{m_1}}{\vec q}'\Big)^2+m_2(2m_1-m_2)}. \label{4point-tree}
\eea
Then, we find that the effective squared mass of the $u$-channel mediator is semi-positive definite for $m_2\leq 2m_1$.  For ${\vec p}={\vec q}'=0$, the above tree-level scattering amplitude diverges at $m_2=2m_1$, which we call the $u$-channel resonance \cite{SRDM}. 
If $\phi_1$ is a stable dark matter, there is no decay width for $\phi_1$, so the above $u$-channel resonance cannot be regularized by a finite decay width, unlike the $s$-channel resonance in which case the mediator has a nonzero width in the region of a resonant enhancement. 
It is crucial to notice that the $u$-channel resonance does not rely on a light mediator, unlike the usual Sommerfeld enhancement from light mediators in the $t$-channels.
In the next subsection, we continue to resum the ladder diagrams shown in the lower Feynman diagram in Fig.~\ref{uFeyn} and obtain the non-perturbative scattering amplitude for the same elastic scattering in the limit of $m_2\sim 2m_1$.

\subsection{Non-perturbative co-scattering}

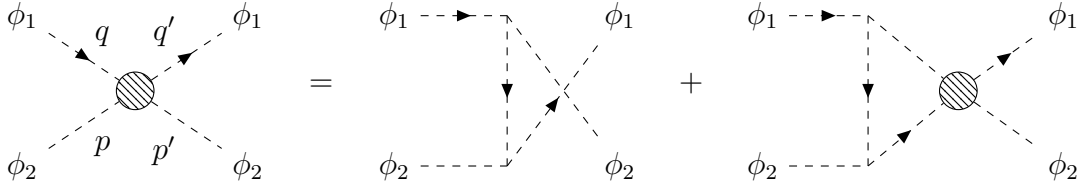
\begin{figure}[!t]
\begin{center}
\begin{equation}
	\begin{tikzpicture}[baseline=(int)]
		\begin{feynman}[small]
			\vertex (ini1) at (-1.5, 1) {\( \phi_1 \)};
			\vertex (ini2) at (-1.5, -1) {\( \phi_2 \)};
			\vertex [blob] (int) at (0, 0) {} ;
			\vertex (f1) at (1.5, 1) {\( \phi_1 \)};
			\vertex (f2) at (1.5, -1) {\( \phi_2 \)};
		
			\diagram* {
				(ini1) -- [charged scalar, edge label = \( q \)] (int) -- [charged scalar, edge label = \( q' \)] (f1),
				(ini2) -- [scalar, edge label' = \( p \)] (int) -- [scalar, edge label' = \( p' \)] (f2),
			};
		\end{feynman}
	\end{tikzpicture}
	\,\,\,\,\, =	\,\,\,\,\,
	\begin{tikzpicture}[baseline=(int)]
		\begin{feynman}[small]
			\vertex (ini1) at (-1.5, 1) {\( \phi_1 \)};
			\vertex (ini2) at (-1.5, -1) {\( \phi_2 \)};
			\vertex (int1) at (0, 1) ;
			\vertex (int2) at (0, -1) ;
			\vertex (f1) at (1.5, 1) {\( \phi_1 \)};
			\vertex (f2) at (1.5, -1) {\( \phi_2 \)};
		
			\diagram* {
				(ini1) -- [charged scalar] (int1) -- [charged scalar] (int2) -- [charged scalar] (f1),
				(ini2) -- [scalar] (int2),
				(int1) -- [scalar] (f2),
				};
		\end{feynman}
	\end{tikzpicture}
		\,\,\,\,\,+	\,\,\,\,\,
	\begin{tikzpicture}[baseline=(int)]
		\begin{feynman}[small]
			\vertex (ini1) at (-2, 1) {\( \phi_1 \)};
			\vertex (ini2) at (-2, -1) {\( \phi_2 \)};
			\vertex (int1) at (-0.6, 1) ;
			\vertex (int2) at (-0.6, -1) ;
			\vertex [blob] (int3) at (0.6, 0) {};
			\vertex (f1) at (2, 1) {\( \phi_1 \)};
			\vertex (f2) at (2, -1) {\( \phi_2 \)};
		
			\diagram* {
				(ini1) -- [charged scalar] (int1) -- [scalar] (int3) -- [charged scalar] (f1),
				(ini2) -- [scalar] (int2) -- [charged scalar] (int3) -- [scalar] (f2),
				(int1) --[charged scalar] (int2),
			};
		\end{feynman}
	\end{tikzpicture} \nonumber 
\end{equation}
\end{center}
\caption{Feynman diagrams for $\phi_1\phi_2\to \phi_1\phi_2$  at non-perturbative level.}
\label{nonpert}
\end{figure}

We present a brief summary of the elastic scattering process, $\phi_1(q)\phi_2(p)\rightarrow \phi_1(q')\phi_2(p')$, at the non-perturbative level, as discussed in Ref.~\cite{SRDM}.

Including the ladder diagrams, we obtain the recursive relation for the non-perturbative four-point function $\Gamma(p,q;p',q')$ for the scattering process, as follows,
\bea
i\Gamma(p,q;p',q')&=&i {\tilde\Gamma}(p,q;p',q') \nonumber \\
&& -\int \frac{d^4k}{(2\pi)^4}\, {\widetilde \Gamma}(p,q;p+q-k,k) G_1(k)G_2(p+q-k) \Gamma(p+q-k,k;p',q')  \nonumber \\
&\approx& -\int \frac{d^4k}{(2\pi)^4}\, {\widetilde \Gamma}(p,q;p+q-k,k) G_1(k)G_2(p+q-k) \Gamma(p+q-k,k;p',q') \label{4point}
\eea
where $G_{1,2}(p)$ are Feynman propagators for $\phi_{1,2}$, given by
\bea
G_{1,2}(p)= \frac{i}{p^2-m^2_{1,2}},
\eea
and $k$ is the loop momentum, and we ignored the perturbative contributions in the approximation. 
Then, defining 
\bea
\chi(p,q;p',q')\equiv G_2(p) G_1(q) \Gamma(p,q; p',q')\equiv \chi(p,q), 
\eea
and multiplying both sides of eq.~(\ref{4point}) by $G_2(p) G_1(q)$,  we rewrite eq.~(\ref{4point}) in momentum space,
\bea
i\chi(p,q)=-G_2(p)G_1(q) \int \frac{d^4 k}{(2\pi)^4} \,  {\widetilde \Gamma}(p,q;p+q-k,k)\, \chi(p+q-k,k). \label{BS0}
\eea

We now make a change of variables by
\bea
P=\frac{1}{2}(p+q),  \quad Q=\mu \Big(\frac{p}{m_2}-\frac{q}{m_1} \Big),
\eea
with $\mu=m_1 m_2/(m_1+m_2)$ being the reduced mass for  the $\phi_1-\phi_2$ system.
Here, $P,  Q$ are proportional to the velocity of the center of mass and the relative velocity, respectively. 
Then, noting 
\bea
\chi(p,q)&=&{\widetilde\chi}(P,Q), \\
\chi(p+q-k,k)&=& {\widetilde\chi}\Big(P,\frac{2\mu}{m_2}P-k\Big).
\eea
where $\widetilde\chi$ is a function of $P$ and $Q$, eq.~(\ref{BS0}) becomes
\bea
i{\widetilde\chi}(P,Q)&=&-G_2\Big(Q+\frac{2\mu}{m_1}\,P\Big) G_1\Big(-Q+\frac{2\mu}{m_2}\, P\Big) \nonumber\\
&&\quad \times  \int \frac{d^4 k'}{(2\pi)^4} \,  {\widetilde \Gamma}(p,q;p+q-k,k)\Big|_{k=-k'+\frac{2\mu}{m_2}P} \, {\widetilde\chi}(P,k')  \label{BS00}
\eea
where a shift in loop momentum \cite{SRDM} is made by $k'=-k+\frac{2\mu}{m_2}P$ and the tree-level four-point function given in eq.~(\ref{4point-tree}) becomes
\bea
 {\widetilde \Gamma}(p,q;p+q-k,k)\Big|_{k=-k'+\frac{2\mu}{m_2}P}
 &=&  \frac{4g^2 m^2_1}{\Big(\sqrt{\frac{m_1}{m_2}} {\vec Q}+\sqrt{\frac{m_2}{m_1}}{\vec k}'\Big)^2+m_2(2m_1-m_2)} \nonumber \\
 &\equiv& U\left(\left|\sqrt{\frac{m_1}{m_2}} {\vec Q}+\sqrt{\frac{m_2}{m_1}}{\vec k}'\right|\right).
\eea

Now we consider the Bethe-Salpeter(BS) equation \cite{SB} for  the $\phi_1-\phi_2$ system from the above results. 
Using the BS wave function in momentum space,
\bea
{\widetilde\psi}_{BS}({\vec Q})= \int \frac{dQ_0}{2\pi} \,{\widetilde\chi}(P,Q),
\eea
and multiplying eq.~(\ref{BS00}) by $\int dQ_0/(2\pi)$, we get
\bea
i{\widetilde\psi}_{BS}({\vec Q})&=& - \int \frac{dQ_0}{2\pi} G_2\Big(Q+\frac{2\mu}{m_1}\,P\Big) G_1\Big(-Q+\frac{2\mu}{m_2}\, P\Big)  \nonumber\\ 
&&\quad\times   \int \frac{d^3 k'}{(2\pi)^3} \,  U\left(\left|\sqrt{\frac{m_1}{m_2}} {\vec Q}+\sqrt{\frac{m_2}{m_1}}{\vec k}'\right|\right)\, {\widetilde\psi}_{BS}({\vec k}').  
\eea
Now making use of the center of mass coordinates for which $P=\frac{1}{2}(m_1+m_2+E,0)$ and $Q=(Q_0,{\vec Q})$ where $E$ is the total kinetic energy, we get \cite{SRDM}
\bea
\left(\frac{{\vec Q}^2}{2\mu}-E \right){\widetilde\psi}_{BS}({\vec Q})=- \int \frac{d^3 k'}{(2\pi)^3} \, {\widetilde V}_1\left(\left|\sqrt{\frac{m_1}{m_2}} {\vec Q}+\sqrt{\frac{m_2}{m_1}}{\vec k}'\right|\right)\, {\widetilde\psi}_{BS}({\vec k}')  \label{BS1}
\eea
with ${\widetilde V}_1=-U/(4m_1m_2)$.
Then, using the BS wave function in position space,
\bea
\psi_{BS}({\vec x})=\int\frac{d^3{\vec Q}}{(2\pi)^3}\, e^{i{\vec Q}\cdot {\vec x}} {\widetilde\psi}_{BS}({\vec Q}),
\eea
 and multiplying both sides of eq.~(\ref{BS1}) by $\int \frac{d^3 {\vec Q}}{(2\pi)^3}\, e^{i{\vec Q}\cdot {\vec x}}$, we get
the BS equation in the following form \cite{SRDM},
\bea
-\frac{1}{2\mu}\,\nabla^2 \psi_{\rm BS}({\vec x}) + V({\vec x})\, \psi_{\rm BS}\Big(-\frac{m_2}{m_1}{\vec x}\Big)= E\psi_{\rm BS}({\vec x})  \label{BS2}
\eea
with
\bea
V({\vec x})=-\frac{\alpha}{r}\, e^{-M r}   \label{Yukawa}
\eea
where $\alpha\equiv\frac{g^2}{4\pi}$ and 
\bea
M\equiv \sqrt{m_2(2m_1-m_2)}\sqrt{\frac{m_2}{m_1}}=m_2\sqrt{2-\frac{m_2}{m_1}}. 
\eea

As a result, we find that the BS equation in eq.~(\ref{BS2}) is a Schr\"odinger-like equation with Yukawa-type potential where the effective mediator mass $M$ vanishes for $m_2=2m_1$ even without a light mediator. 
Moreover, we find that the argument of the wave function in the potential term in eq.~(\ref{BS2}) is non-standard, showing the delayed interactions with $u$-channel resonances in general as will be discussed later in more detail.
This result is due to the fact that a nonzero energy exchange discussed below eq.~(\ref{energytransfer}) contributes to the effective mass for the exchanged particle in the $u$-channel process and makes the $u$-channel interaction necessarily non-instantaneous for the $u$-channel resonance. Then, we need an early history of the wave function in the position space in order to solve the above BS equation.

\subsection{Solutions to the Bethe-Salpeter equation}

As the potential in the BS equation is central, we can make a separation of variables  for the BS wavefunction as
\bea
\psi_{BS}({\vec x})= R_{l}(r) Y_l^m(\theta,\phi). 
\eea
Then, we also get
\bea
\psi_{\rm BS}\Big(-\frac{m_2}{m_1}{\vec x}\Big)&=&  R_{l}\Big(\frac{m_2}{m_1}r\Big)Y_l^m(\pi-\theta,\phi+\pi) \nonumber \\
&=& (-1)^l R_{l}\Big(\frac{m_2}{m_1}r\Big) Y_l^m(\theta,\phi). 
\eea
As a result, the BS equation (\ref{BS2}) leads to the following radial equation, 
\bea
\Bigg[-\frac{1}{2\mu}\bigg(\frac{d^2}{dr^2}+\frac{2}{r}\,\frac{d}{dr} \bigg)+\frac{l(l+1)}{2\mu r^2}\Bigg] R_{l}(r)-\frac{\alpha}{r}  \, e^{-Mr}\, (-1)^l\, R_{l}(b r)= E\, R_{l}(r).  \label{Scheq0}
\eea
with 
\bea
b=\frac{m_2}{m_1}.
\eea
Therefore, we have got striking results in the radial equation. The $u$-channel interaction is attractive for even $l$, whereas it  becomes repulsive for odd $l$. This is a remarkable result. The $u$-channel interaction interchanges between $\phi_1$ and $\phi_2$ appearing in the BS wave function for the non-perturbative scattering amplitude, in contrast to the case for the $t$-channel interaction. Then,  for odd $l$, the BS wave function for the $\phi_1-\phi_2$ system is  anti-symmetric, so the potential energy in the position space flips in sign. 

With a change of variables, $x=\frac{1}{2}\mu \alpha r$, the above equation becomes
\bea
\Bigg[\frac{d^2}{dx^2}+\frac{2}{x}\frac{d}{dx} -\frac{l(l+1)}{x^2}\Bigg] R_l(x)+\frac{4}{x} \, e^{-c\,x}\, (-1)^l\, R_l(bx)+a^2\, R_l(x)=0  \label{Scheq01}
\eea
with
\bea
a=\frac{2v}{\alpha},\quad c=\frac{2M}{\mu\alpha}.
\eea
Here, $v$ is the relative velocity between $\phi_1$ and $\phi_2$. 

Furthermore, by writing 
\bea
R_{l}(x)= \frac{u_l(x)}{x},
\eea
the Schro\"odinger-like equation in eq.~(\ref{Scheq01}) is reduced to a simpler form, as follows,
\bea
\Bigg[\frac{d^2}{dx^2} -\frac{l(l+1)}{x^2}\Bigg] u_l(x)+\frac{4}{b\, x} \, e^{-c\,x}\, (-1)^l\, u_l(bx)+a^2\, u_l(x)=0. \label{xeq}
\eea
Then, with a change of variables, $x=e^{-\rho}$, the above equations can be rewritten as a delay differential equation with constant shift, $\ln b$, 
\bea
\frac{d^2}{d\rho^2} {\tilde u}_l(\rho)+\frac{d}{d\rho}{\tilde u}_l(\rho)-l(l+1) {\tilde u}_l(\rho)+\frac{4}{b}\, {\rm exp}\Big[{-\rho-c\, e^{-\rho}}\Big]\, {\tilde u}_l(\rho-\ln b)+a^2e^{-2\rho}\, {\tilde u}_l(\rho)=0.
\eea
with ${\tilde u}_l(\rho)=u_l(e^{-\rho})$. 

We remark on the non-linear behavior of the obtained differential equation in eq.~(\ref{xeq}).  
For the bound-state solution to exist, we need to choose $u_l\sim x^{l+1}$ for $x\rightarrow 0$ and $u_l\sim e^{-\kappa x}$ with $\kappa=|a|$  for $x\rightarrow \infty$. 
Then, we can choose the solution in the following form,
\bea
u_l(x)= x^{l+1} e^{-\kappa x} f(x). 
\eea
But, plugging this solution into eq.~(\ref{xeq}), the equation cannot be cast into a linear differential equation for $f(x)$, because of $u_l(bx)=(bx)^{l+1} e^{-b\kappa x} f(bx)$.  The contribution from the potential term at large distances can be ignored such that $u_l\sim e^{-\kappa x}$ is a correct asymptotic solution.
However,  a non-linear term $u_l(bx)$ changes the form of the solution in the intermediate region.

\subsection{Boundary conditions}

For elastic scattering process for two-component dark matter, it is necessary to impose the boundary conditions for the BS wave function at the origin and a far distance from the collision point. We first discuss the case for the $s$-wave scattering and then generalize it to the case for higher order partial waves.

For an unbounded $s$-wave scattering process, we can expand the radial wave-function as $R_0(x)=R_0(0)+x R'_0(0)+\cdots$  near $x=0$, so we can read off the radial solution at $x=0$ by  $R_0(0)= A\neq 0$.  Moreover, we also choose the normalization of the plane-wave limit for the radial solution at $x=\infty$, as follows,
\bea
R_0(x)\longrightarrow \frac{\sin(a x+\delta_0)}{ax}. \label{bc2}
\eea
As a result, the boundary conditions for the radial wave-function in the new coordinate $\rho$ become
\bea
{\tilde u}_0(\rho)& \longrightarrow&\frac{1}{a}\, \sin(a\, e^{-\rho}+\delta_0), \qquad \rho \to -\infty,  \label{ubc1} \\
{\tilde u}_0(\rho)&\longrightarrow& A\, e^{-\rho}, \,\,\,\,\,\quad\qquad\qquad\quad \rho\to +\infty \label{ubc2} 
\eea
where $\delta_0$ is the phase shift for $s$-wave.
Then, after imposing the first boundary condition in eq.~(\ref{ubc1}) as a history function at $\rho=-\infty$ for the  delay differential equation, from the second boundary condition in eq.~(\ref{ubc2}), we can determine the phase shift $\delta_0$ as well as the Sommerfeld factor, $S_0=A^2$.

For a nonzero $l$, the radial wave-function goes by $R_l(r)\to A_l\,  (k r)^l=A_l (a x)^l$  with $k=\mu v$, for $x\to 0$, and the plane-wave limit for the radial solution at $x=\infty$ should be taken to
\bea
R_l(x)\longrightarrow  e^{i\delta_l}\,\cdot\frac{\sin\big(a x+\delta_l-\frac{l\pi}{2}\big)}{ax} \label{bc3}
\eea
where $\delta_l$ are  the phase shifts for higher partial waves.
Then, the boundary conditions in the new coordinate  $\rho$ become
\bea
{\tilde u}_l(\rho)& \longrightarrow&\frac{ e^{i\delta_l}}{a}\, \sin\Big(a\, e^{-\rho}+\delta_l-\frac{l\pi}{2}\Big), \qquad \rho \to -\infty,  \label{ubc1n} \\
{\tilde u}_l(\rho)&\longrightarrow& A_l\,  a^l\, e^{-(l+1)\rho}, \quad\qquad\,\,\, \rho\to +\infty .\label{ubc2n} 
\eea
For a complex constant, $A_l=c_l+i d_l$, the above boundary conditions become
\bea
{\rm Re}[{\tilde u}_l(\rho)]& \longrightarrow&\frac{\cos\delta_l}{a} \,  \sin\Big(a\, e^{-\rho}+\delta_l-\frac{l\pi}{2}\Big), \qquad \rho \to -\infty,  \label{ubc1na} \\
{\rm Re}[{\tilde u}_l(\rho)]&\longrightarrow& c_l \,a^l \,e^{-(l+1)\rho},  \quad\qquad\,\,\, \rho \to +\infty,  \label{ubc1na} 
\eea
whereas the imaginary part of ${\tilde u}_l(\rho)$ is not independent, but it depends on the same phase shift $\delta_l$ appearing in the boundary condition in the plane-wave limit, as follows,
\bea
{\rm Im}[{\tilde u}_l(\rho)]& \longrightarrow&\frac{\sin\delta_l}{a} \,  \sin\Big(a\, e^{-\rho}+\delta_l-\frac{l\pi}{2}\Big), \qquad \rho \to -\infty,  \label{ubc1nb} \\
{\rm Im}[{\tilde u}_l(\rho)]&\longrightarrow& d_l \,a^l \,e^{-(l+1)\rho},  \quad\qquad\,\,\, \rho \to +\infty. \label{ubc1nb} 
\eea
In this case, we can determine the Sommerfeld factor by the following differentiation of the radial wave-function,
\bea
S_l &=&  \bigg|  \frac{(2l+1)!!}{k^l l!} \,\frac{\partial^l R_l(r)}{\partial r^l}\bigg|_{r=0}\bigg|^2 \nonumber \\
&=& \Big|(2l+1)!! \, A_l \Big|^2 \nonumber \\
&=&((2l+1)!!)^2 (c^2_l + d^2_l).
\eea
Therefore, for higher order partial waves, the BS wave function at the origin takes a complex value, in terms of which the Sommerfeld factor can be determined. 
We note that it was shown in Ref.~\cite{yu} that the spherical Bessel and Neumann functions can be taken at large radii for a better convergence of the numerical solutions instead of the sinusoidal functions, but it is enough for us to take the above simple boundary conditions  for keeping the leading order effects of the $s$-wave scattering in the later discussion.

First, taking the Coulomb limit with $M=0$ (or $m_2=2m_1$) and choosing the $s$-wave solution with $l=0$, from eq.~(\ref{xeq}),  
we obtain
\bea
\frac{d^2 u_0}{dx^2} + \frac{2}{x}\,u_0(2x) + a^2 u_0(x)=0. \label{swave2}
\eea
Then, in the new coordinate $\rho$, the above equations can be rewritten as a delay differential equations with constant shift, $\ln 2$, 
\bea
\frac{d^2}{d\rho^2} {\tilde u}_0(\rho)+\frac{d}{d\rho}{\tilde u}_0(\rho)+2 e^{-\rho}\, {\tilde u}_0(\rho-\ln 2)+a^2e^{-2\rho}\, {\tilde u}_0(\rho)=0
\eea

Next, for more general cases with $M\neq 0$ (or $m_2<2m_1$), we consider the equation for  the $s$-wave radial wave-function as
\bea
\frac{d^2u_0}{dx^2} +\frac{4}{b\, x} \, e^{-c\,x}\,  u_0(bx)+a^2\, u_0(x)=0.
\eea
In this case, the corresponding delay differential equations are
\bea
\frac{d^2}{d\rho^2} {\tilde u}_0(\rho)+\frac{d}{d\rho}{\tilde u}_0(\rho)+\frac{4}{b}\, {\rm exp}\Big[{-\rho-c\, e^{-\rho}}\Big]\, {\tilde u}_0(\rho-\ln b)+a^2e^{-2\rho}\, {\tilde u}_0(\rho)=0.
\eea
Thus, the delay differential equations still have constant shifts in the potential terms for $m_2<2m_1$. 
Therefore, we can impose the same boundary conditions at $\rho\to \pm \infty$  as those in eqs.~(\ref{ubc1}) and (\ref{ubc2}).

For $M=0$ and nonzero $l$, for $\alpha>0$, the effective potential is attractive only for even $l$, so there is a Sommerfeld enhancement factor in this case. If $\alpha<0$, which is the case for other dark matter candidates in the later section, the effective potential is attractive only for odd $l$, so the Sommerfeld enhancement is relevant. 
For $M\neq 0$ and nonzero $l$, we can make similar discussions for the Sommerfeld factor as in the case with $M=0$, but with a Yukawa-like potential.

\section{Non-perturbative effects from $u$-channel resonances}

We discuss the general consequences of the non-perturbative effects from $u$-channel resonances for $2\to 2$ co-scattering, $2\to 2$ and $3\to 2$ semi-annihilation processes.

\subsection{$2\to 2$ co-scattering}

Taking $m_2\lesssim 2m_1$ for which the effective mass $M$ to be positive, 
we have the approximate Coulomb limit for $\phi_1\phi_2\to \phi_1\phi_2$ due to the $u$-channel resonance. 
In this case, since  the heavier particle $\phi_2$ is stable,  two components of dark matter, $\phi_1$ and $\phi_2$, coexist at present, so they undergo the elastic scattering. 

The elastic scattering cross section for $\phi_1\phi_2\to \phi_1\phi_2$ can be enhanced by the non-perturbative effect due to the effective light mediator in the $u$-channel.
For  $s$-wave dominance, the corresponding total scattering cross section is given by
$\sigma_{\rm self}= \frac{4\pi}{k^2}\, \sin^2\delta_0$.
Therefore, we can achieve a large self-scattering cross section for dark matter up to unitarity bound, thanks to the $u$-channel resonance. In this case, the self-scattering cross section is enhanced at small velocities to solve the small-scale problems at galaxies and the diversity problem \cite{sidm}, while being consistent with the bounds from galaxy clusters \cite{smallscale3}.

In practice, the energy density fractions of two dark matter components can be different at present.
So, assuming that  $\phi_1\phi_2\to \phi_1\phi_2$ is a dominant process for self-scattering, we define the effective self-scattering rate for dark matter, required to solve the small-scale problems \cite{SRDM}, as follows,
\bea
\Gamma_{\rm scatt}&=& \frac{n_1 n_2}{n^2_{\rm DM}}\, \sigma_{\rm self}\, n_{\rm DM}\,v \nonumber \\
&=& \rho_{\rm DM}\,\cdot\frac{r_1(1-r_1)\sigma_{\rm self}v}{m_2 \,r_1+m_1(1-r_1)} \equiv \frac{\rho_{\rm DM} \sigma_{\rm self}v}{m_{\rm eff}} \label{scattrate}
\eea
where $r_1=\Omega_1/\Omega_{\rm DM}$ is the fraction of the relic density for the $\phi_1$ particle, $\rho_{\rm DM}$ is the local dark matter (DM) energy density at galaxies, and $m_{\rm eff}=(m_2 \,r_1+m_1(1-r_1))/(r_1(1-r_1))$. Then, in order to solve the small-scale problems at galaxies, we would need $\sigma_{\rm self}/m_{\rm eff}=0.1-10\,{\rm cm^2/g}$ \cite{sidm0,smallscale3,diversity,sidm}.

We note that there is another elastic scattering, $\phi_1\phi^{(*)}_1\rightarrow \phi_1\phi^{(*)}_1$ and its complex conjugate, which can have a similar resonance enhancement near $m_2\sim 2m_1$. 
But, for $m_2\lesssim 2m_1$ as in the case of $u$-channel enhancement, the s-channel enhancement is less significant, because the center of mass energy for a pair of $\phi_1$'s is always greater than $m_2$.
Therefore, the non-perturbative effects for  $\phi^{(*)}_1\phi_2\to \phi^{(*)}_1\phi_2$ give rise to dominant effects for rendering dark matter self-interacting.

\subsection{$2\to 2$ semi-annihilations}

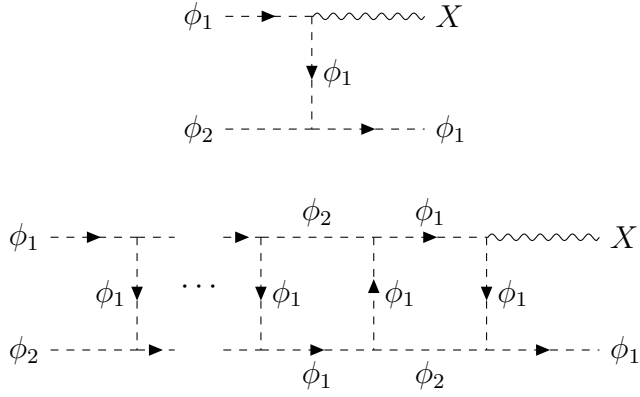
\begin{figure}[!t]
\begin{center}
\begin{tikzpicture} [baseline=($0.5*(ini1)+0.5*(ini2)$)]
	\begin{feynman} [inline=($0.5*(ini1)+0.5*(ini2)$), medium]
		\vertex (ini1) { \( \phi_1 \) } 	;	\vertex [below = of ini1] (ini2) { \( \phi_2 \) } ;
		\vertex [right = of ini1](int1) ;	\vertex [right = of ini2](int2) ;
		\vertex [right = of int1](f1) { \( X \) } ;	\vertex [right = of int2](f2) { \( \phi_1 \) } ;
		
		\diagram*{
			(ini1) -- [charged scalar] (int1) -- [photon] (f1),
			(ini2) -- [scalar] (int2) -- [charged scalar] (f2),
			(int1) -- [charged scalar, edge label = \( \phi_1 \)] (int2),
		};
	\end{feynman}
\end{tikzpicture} 
\be
\begin{tikzpicture} [baseline=($0.5*(ini1)+0.5*(ini2)$)]
	\begin{feynman} [inline=($0.5*(ini1)+0.5*(ini2)$), medium]
		\vertex (ini1) { \( \phi_1 \) } ;
		\vertex [below = of ini1] (ini2) { \( \phi_2 \) } ;
		\vertex [right = of ini1](int1) ; \vertex [right = of ini2](int2) ;
		\vertex [right = 0.5cm of int1](f1) ; \vertex [right = 0.5cm of int2](f2) ;
		
		\diagram*{
			(ini1) -- [charged scalar] (int1) -- [scalar] (f1),
			(ini2) -- [scalar] (int2) -- [charged scalar] (f2),
			(int1) -- [charged scalar, edge label' = \( \phi_1 \)] (int2),
		};
	\end{feynman}
	\end{tikzpicture}
	\cdots
	\begin{tikzpicture} [baseline=($0.5*(ini1)+0.5*(ini2)$)]
	\begin{feynman} [inline=($0.5*(ini1)+0.5*(ini2)$), medium]
		\vertex (ini1) ;
		\vertex [below = of ini1] (ini2) ;
		\vertex [right =0.5cm of ini1] (u1) ; \vertex [right = of u1] (u2) ; \vertex [right = of u2] (u3) ;  \vertex [right = of u3] (f1) { \( X \) } ;
		\vertex [right = 0.5cm of ini2] (d1) ; \vertex [right = of d1] (d2) ; \vertex [right = of d2] (d3) ;  \vertex [right = of d3] (f2) { \( \phi_1 \) } ;
				
		\diagram*{
			(ini1) -- [charged scalar] (u1) -- [scalar, edge label = \( \phi_2 \)] (u2) -- [charged scalar, edge label = \( \phi_1 \)] (u3) -- [photon] (f1),
			(ini2) -- [scalar] (d1) -- [charged scalar, edge label' = \( \phi_1 \)] (d2) -- [scalar, edge label' = \( \phi_2 \)] (d3) -- [charged scalar] (f2),
			(u1) -- [charged scalar, edge label = \( \phi_1 \)] (d1), (u2) -- [anti charged scalar, edge label = \( \phi_1 \)] (d2), (u3) -- [charged scalar, edge label = \( \phi_1 \)] (d3), 
		};
	\end{feynman}
	\end{tikzpicture}
	\nonumber
\ee	
\end{center}
\caption{Feynman diagrams for $2\to 2$ dark matter annihilation, $\phi_1\phi_2\to \phi_1 X$, at the tree level (above), and the corresponding ladder diagrams (below).}
\label{22ann0}
\end{figure}

As shown in Fig.~\ref{22ann0}, the annihilation cross section for $\phi_1\phi_2\to \phi_1 X$ with $X$ being an extra mediator  can be enhanced by the Sommerfeld factor coming from the ladder diagrams with the $u$-channel resonance.

Taking into account the tree-level and ladder diagrams for $\phi_1\phi_2\to \phi_1 X$ in Fig.~\ref{22ann0}, the corresponding non-perturbative scattering amplitude  can be written in terms of the one without ladder diagrams \cite{cassel}, as follows,
\bea
{\cal M}_{2\to2, w}(p_1,p_2;\{p_f\})=\int\frac{d^4 k}{(2\pi)^4}  \, {\cal M}_{2\to 2,w/o} (p_3,p_4;\{p_f\}) G_2(p_3) G_1(p_4) \Gamma(p_1,p_2;p_3,p_4)
\eea
where ${\cal M}_{2\to 2,w/o} (p_3,p_4;\{p_f\})$ is the scattering amplitude without ladder diagrams and $\{p_f\}$ are the final momenta.
Then, in the non-relativistic limit, the above scattering amplitude can be approximated in the center of mass frame, as follows,
\bea
{\cal M}_{2\to2, w}(p_1,p_2;\{p_f\})\simeq \int \frac{d^3 k}{(2\pi)^3}\, {\cal M}_{2\to 2,w/o}({\vec k},-{\vec k};\{p_f\})\,  {\widetilde \psi}^*_{BS}({\vec k}).
\eea

As a result, Taking a partial-wave expansion for the scattering amplitude for $\phi_1\phi_2\to \phi_1 X$ at tree level to be ${\cal M}_{2\to 2,w/o}({\vec p},-{\vec p};\{p_f\})=a_l |{\vec p}|^l Y_{lm}(\Omega_p)$  in the center of mass frame, and using the BS wave function in the momentum space, $ {\widetilde \psi}_{BS}({\vec k})= \sum_{l',m'}{\widetilde R}_{l'}(k) Y_{l'm'}(\Omega_k)$,  we obtain the non-perturbative scattering amplitude as
\bea
{\cal M}_{2\to2, w}&=& a_l \int \frac{dk}{(2\pi)^3} \, k^{l+2} {\tilde R}_l(k) \int d\Omega_k \sum_{l',m'} Y_{lm}(\Omega_k) Y^*_{l'm'}(\Omega_k) Y_{l'm'}(\Omega_p) \nonumber \\
&=& a_l \sum_{l',m'} \int\frac{dk}{2\pi} \, k^{l+2}\, {\widetilde R}_l(k)\,  \delta_{ll'} \delta_{mm'} Y_{l'm'}(\Omega_p).
\eea 
Consequently, similarly to the standard case for $2\to 2$ annihilations \cite{cassel}, the Sommerfeld factor for $\phi_1\phi_2\to \phi_1 X$ becomes
\bea
S_l&=& \frac{\big|{\cal M}_{2\to2, w}\big|^2}{\big|{\cal M}_{2\to2, w/o}\big|^2} \nonumber \\
&=& \bigg| \int \frac{dk}{2\pi^2} \,k^{l+2} \,  {\widetilde R}_l(k) \, |p_1|^{-l} \bigg|^2 \nonumber \\
&=& \bigg|  \frac{(2l+1)!!}{|{\vec p}_1|^l l!} \,\frac{\partial^l R_l(r)}{\partial r^l}\bigg|_{r=0}\bigg|^2.
\eea

Then, the non-perturbative cross section for $\phi_1\phi_2\to \phi_1 X$ becomes
\bea
(\sigma v)_{2\to 2, w}=S_l\, \cdot (\sigma v)_{2\to 2, w/o}
\eea
where the $2\to 2$ semi-annihilation cross section without ladder diagrams is
\bea
(\sigma v)_{2\to 2, w/o}=\frac{|{\cal M}_{2\to2, w/o}|^2}{32\pi m_1 m_2}\,  \sqrt{1-\frac{(m_1+m_X)^2}{(m_1+m_2)^2}}\,\sqrt{1-\frac{(m_1-m_X)^2}{(m_1+m_2)^2}}.
\eea

Imposing the unitarity bound on the $s$-wave annihilation with $ (\sigma v)_{2\to 2, w/o}=\alpha^2_{\rm eff}/m^2_1$, we set the bound on the Sommerfeld factor by $S_0<4\pi  v m^2_1/(\alpha^2_{\rm eff} k^2)$, which becomes
$S_0<4\times 10^4(0.6/v)(\frac{1}{30}/\alpha_{\rm eff})^2$ for $m_2\simeq 2m_1$. Thus,  as will be discussed in the later section, $\phi_1\phi_2\to \phi_1 X$ could be constrained by CMB or indirect detection experiments from cosmic rays, depending on the Sommerfeld factor and the decay products of $X$. 

If $\phi_1\phi_2\to \phi_1 X$ is dominated by $p$-wave at tree level, there is no Sommerfeld enhancement due to the $u$-channel resonance, because the effective potential becomes repulsive for $p$-wave as discussed below eq.~(\ref{xeq}) and higher order partial waves are velocity-suppressed.

\subsection{$3\rightarrow 2$ semi-annihilations}

\begin{figure}[!t]
\begin{center}
\begin{tikzpicture} 
	\begin{feynman}
		\vertex (ini1) { \( \phi_1 \) } ;
		\vertex [below = of ini1] (ini2) { \( \phi_1 \) } ;
		\vertex [below = of ini2] (ini3) { \( \phi_1 \) } ;
		\vertex [below right = of ini1] (int1a) ;
		\vertex [below = 1cm of int1a] (int1b) ;
		\vertex [right = of int1a] (int2) ;
		\vertex [above right = of int2] (f1) { \( \phi_1 \) } ;
		\vertex [below right = of int2] (f2) { \( \phi_2 \) };
		
		\diagram*{
			(ini1) -- [charged scalar] (int1a) -- [charged scalar, edge label = \( \phi_1 \)] (int2),
			(f1) -- [anti charged scalar] (int2) -- [scalar] (f2),
			(ini2) -- [charged scalar] (int1b) -- [charged scalar] (ini3),
			(int1a) -- [scalar, edge label = \( \phi_2 \)] (int1b), 
		};
	\end{feynman}
\end{tikzpicture}
\end{center}
\caption{Feynman diagram for $3\to 2$ dark matter annihilation, $\phi_1\phi_1\phi^*_1\to \phi_1 \phi_2$.}
\label{32ann}
\end{figure}
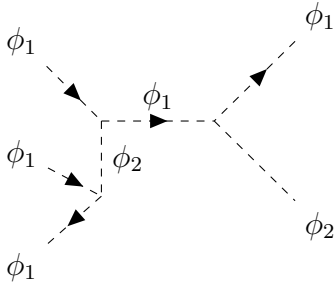

The $u$-channel resonance also appears as a part of the Feynman diagram for higher order annihilation processes such as $3\to 2$. In the presence of a large self-interaction for dark matter, the $3\to 2$ annihilation processes can account for the freeze-out for Strongly Interacting Massive Particles (SIMPs) \cite{simp,simp2,axionportal,simpz3,simprelic,review,vsimp1,vsimp2}.
Due to the triple interaction for dark matter in our model,  a $3\to 2$ semi-annihilation channel for dark matter, $\phi_1\phi_1\phi^*_1\to \phi_1\phi_2$, is allowed for $2m_1>m_2$. But, there is no Sommerfeld enhancement for the initial state, unlike the $2\to 2$ annihilation. Nonetheless, the $u$-channel resonance in the $3\to 2$ annihilation process still enhances the cross section.

For  $\phi_1\phi_1\phi^*_1\to \phi_1\phi_2$, in the limit of vanishing velocities for dark matter, the $3\to 2$ semi-annihilation cross section at tree level \cite{review} is given by
\bea
(\sigma v^2_{\rm rel})_{\phi_1\phi_1\phi^*_1\to \phi_1\phi_2} = \frac{|{\cal M}_{\phi_1\phi_1\phi^*_1\to \phi_1\phi_2}|^2}{72 \pi m^3_1}\, \sqrt{\bigg(1-\frac{m^2_2}{4m^2_1}\bigg)\bigg(1-\frac{m^2_2}{16m^2_1}\bigg)} \label{32tree}
\eea
where the squared scattering amplitude at tree level is given by
\bea
|{\cal M}_{\phi_1\phi_1\phi^*_1\to \phi_1\phi_2}|^2= \frac{288 g^6 m^6_1}{(4m^2_1-m^2_2)^4}. 
\eea
Here, there is a singularity at $m_2=2m_1$, due to the combination of the $s$-channel resonance with $\phi_2$ and the $u$-channel resonance with $\phi_1$.
Then, for $m_2=2m_1(1-\Delta)$ with $0<\Delta\ll 1$, we can approximate the phase space factor to get $ (\sigma v)_{\phi_1\phi_1\phi^*_1\to \phi_1\phi_2}$ as
\bea
(\sigma v^2_{\rm rel})_{\phi_1\phi_1\phi^*_1\to \phi_1\phi_2} \simeq \frac{\sqrt{6}\pi^2\alpha^3}{32 m^5_1}\, \Delta^{-7/2}. 
\eea
Therefore, for a small $\Delta>0$,  the $3\to 2$ semi-annihilation cross section can be enhanced through the $u$-channel resonance \cite{vsimp1}, so it is free of the perturbativity problem in SIMP dark matter models \cite{resonance,ko}.  
For a finite temperature, the resonance poles in the $3\to 2$ annihilation are regularized by a nonzero velocity of dark matter \cite{vsimp1}, so the enhancement gets less significant.  Thus, for the freeze-out of WIMP-like dark matter, the $3\to 2$ annihilation processes are sub-dominant as compared to the $2\to 2$ annihilation processes.

We note that the $3\to 2$ annihilation cross section is subject to the perturbativity unitarity bound \cite{32signals}, as follows,
\bea
\langle(\sigma v^2_{\rm rel})_{\phi_1\phi_1\phi^*_1\to \phi_1\phi_2} \rangle\leq \frac{96\sqrt{3}\pi^2 x^2}{m^5_1}\, \frac{g_{\phi_2}}{g_{\phi_1}^2} 
\eea 
where $x=m_1/T$ and $g_{\phi_1}, g_{\phi_2}$ are the degrees of freedom of $\phi_1$ and $\phi_2$, respectively, given by $g_{\phi_1}=2$ and $g_{\phi_2}=1$ in our model.

\section{Effective theory for self-resonant dark matter}

In this section, we consider the effective theory for self-resonance dark matter with two components.  
We take various combinations of scalar, fermion and vector dark matter with minimal self-interactions between them.  
The results in this section are summarized in Table~\ref{table:srdm}.

 \begin{table}[h]

\begin{center}
 {\scriptsize
\begin{tabular}{c||ccccc}
\hline
\hline
 {\rm dark matter}  & {\rm scalar }  & {\rm pseudo-scalar}  & {\rm fermion}  & {\rm vector} & {\rm axial-vector}  \\ 
\hline 
\hline
\\
{\rm scalar}($\phi$) &  $+4g^2 m^2_\phi$  & $+4g^2 m^2_\phi$ &  $\pm 2y^2_\chi m_\chi (2m_\chi-m_\phi)$  & NA & NA \\ 
\\
{\rm pseudo-scalar}($a$)&  $-$  &  $-$ &  $\mp 2\lambda^2_\chi m_\chi m_a$   & NA & NA \\
\\
{\rm fermion}($\chi$)  &  $-$  & $-$  & NA   & $\mp 2g^2_{Z'} m_\chi m_{Z'}$ & $\pm 2g^2_{A'} m_\chi (2m_\chi-m_{A'})$ \\ 
\\
{\rm vector}($Z'$)  & NA  & NA & $-$  & $-6g^2_X m_X(2m_X-m_{X_3})$ & NA \\ 
\\
{\rm axial-vector}($A'$) & NA   & NA  &  $-$  &  $-$  & NA  \\ 
\\
\hline
\hline
\end{tabular}
}
\caption{Tree-level amplitudes for elastic co-scattering for multi-component dark matter with minimal couplings, divided by the $u$-channel propagator.  When there are two signs, the upper (lower) sign denotes the fermion (anti-fermion) scattering. ``NA'' implies that the processes are disallowed or velocity-suppressed. For vector ($X$)-vector ($X_3$) scattering, we consider gauge bosons in a non-abelian gauge theory in the dark sector. Here, $X$ is a complex gauge boson and $X_3$ is a real gauge boson. We remark that the overall positive sign implies an attractive Yukawa potential for the $s$-wave scattering. } 
\label{table:srdm}
\end{center}
\end{table}

We first discuss the combination of scalar dark matter with two more scalar particles, one vector dark matter or one pseudo-scalar dark matter. Then, we also take the combination of fermion dark matter with one scalar dark matter, pseudo-scalar dark matter, vector dark matter or axial-vector dark matter. Finally, we also mention the case for two-component vector dark matter from a non-abelian gauge group.

\subsection{Scalar dark matter}

\underline{Case with two scalar components}

In scalar dark matter models in eq.~(\ref{scalardm}), we can introduce an extra scalar $\phi_3$ with mass $m_3$ and the following self-interaction Lagrangian,
\bea
{\cal L}_{\rm int}=-2g \,m_1 \phi_1\phi_2\phi^*_3+{\rm h.c.}.
\eea
Then, $\phi_1\phi_2\rightarrow \phi_1\phi_2$ and its complex conjugate can be mediated by an additional scalar $\phi_3$ with mass $m_3$ in the $u$-channel, so the corresponding tree-level scattering amplitude is given by
\bea
 {\widetilde \Gamma}(p,q;p',q') \approx \frac{4g^2 m^2_1}{\Big(\sqrt{\frac{m_1}{m_2}} {\vec p}-\sqrt{\frac{m_2}{m_1}}{\vec q}'\Big)^2+m^2_3-(m_1-m_2)^2}. \label{4point-tree}
\eea
Then, for $m_3=|m_1-m_2|$, there is an $u$-channel enhancement of the scattering amplitude too. 
In this case, the effective mediator mass in the Yukawa potential in eq.~(\ref{Yukawa}) becomes
\bea
M&=&\sqrt{\frac{m_2}{m_1}}\, \sqrt{m^2_3-(m_1-m_2)^2} \nonumber \\
&=&\sqrt{\frac{m_2}{m_1}}\, \sqrt{(m_3+m_1-m_2)(m_3+m_2-m_1)}. 
\eea

Similarly,  $\phi_3\phi_2\rightarrow \phi_3\phi_2$ and its complex conjugate can be enhanced through the $u$-channel resonance. The corresponding tree-level scattering amplitude is given by
\bea
 {\widetilde \Gamma}(p,q;p',q') \approx \frac{4g^2 m^2_1}{\Big(\sqrt{\frac{m_3}{m_2}} {\vec p}-\sqrt{\frac{m_2}{m_3}}{\vec q}'\Big)^2+m^2_1-(m_3-m_2)^2}. \label{4point-tree}
\eea
In this case, we also identify  the effective mediator mass in the Yukawa potential in eq.~(\ref{Yukawa})  as
\bea
M&=&\sqrt{\frac{m_2}{m_3}}\, \sqrt{m^2_1-(m_3-m_2)^2}  \nonumber \\
&=&\sqrt{\frac{m_2}{m_3}}\,\sqrt{(m_1+m_3-m_2)(m_1+m_2-m_3)}.
\eea
Then, for $m_1=|m_3-m_2|$, there is an $u$-channel enhancement of the scattering amplitude. 
For $m_3=m_1$, we can recover the case with two-component scalar dark matter, for which the $u$-channel resonance condition becomes $m_2=2m_1$.

\underline{Case with vector dark matter}

We also consider a complex scalar dark matter $\phi$ coupled to a massive vector dark matter $Z'$, with the following interaction,
\bea
{\cal L}_{\rm int}= ig_{Z'} Z'_\mu (\phi \partial^\mu \phi^* - \phi^* \partial^\mu\phi).
\eea
We note that there are also $|\phi|^2 Z'_\mu Z^{\prime \mu}$ contact interactions for scalar dark matter, but they are not relevant for the $u$-channel resonance, so we don't consider them. 
In this case, the tree-level scattering amplitude for $\phi(q)Z'(p)\to \phi(q') Z'(p')$ is given by
\bea
 {\widetilde \Gamma}(p,q;p',q') =\frac{-ig^2_{Z'}}{ (q-p')^2-m^2_\phi}\,\cdot \epsilon^\mu(p)\epsilon^{*\nu}(p') (2q-p')_\nu (q-p'+q')_\mu.
\eea
Thus, in the non-relativistic limit, we get the polarization vectors for $Z'$ as $\epsilon^\mu(p)\simeq \delta^{\mu i}$ and  $\epsilon^{*\nu}(p)\simeq \delta^{\nu j}$, with $i,j=1,2,3$, so the above scattering amplitude is velocity-suppressed. Then, this case is not relevant for self-resonant dark matter. 

We can also consider a scalar dark matter $\phi$ plus a pseudo-scalar dark matter $a$ with the interaction Lagrangian, ${\cal L}_{\rm int}=-2gm_\phi \phi \, a^2$. In this case, the elastic scattering, $\phi \,a\to \phi \, a$, can be enhanced at a small velocity due to the $u$-channel resonance at $m_a=2m_\phi$, so this case is similar to the case for two scalar component dark matter, as discussed for $\phi_1$ and $\phi_2$ in the previous section.
Moreover, the option with scalar dark matter and fermion dark matter will be discussed in the next subsection.

\subsection{Fermion dark matter}

\underline{Case with scalar dark matter}

We consider a two-component dark matter with  fermion dark matter $\chi$ and scalar dark matter $\phi$, with the following Lagrangian,
\bea
{\cal L}_\chi= \frac{1}{2}(\partial_\mu\phi)^2-\frac{1}{2} m^2_\phi \phi^2+{\bar\chi}(i\gamma^\mu\partial_\mu-m_\chi)\chi - y_\chi \phi {\bar\chi} \chi
\eea
where $y_\chi$ is the Yukawa coupling for fermion dark matter.
Then, we can consider a similar elastic scattering process, $\chi \phi\to \chi\phi$, and its complex conjugate process, and  fermion dark matter $\chi$ can contribute to the $u$-channel resonance.

For elastic scattering processes, $\chi(q)\phi(p)\to \chi(q')\phi(p')$ and  ${\bar\chi}(q)\phi(p)\to {\bar\chi}(q')\phi(p')$, we get the scattering amplitude at tree level in the non-relativistic limit, respectively, as
\bea
{\tilde\Gamma}_\chi(p,q;p',q') = \frac{\pm 2y^2_\chi m_\chi(2m_\chi-m_\phi)}{\Big(\sqrt{\frac{m_\chi}{m_\phi}} {\vec p}-\sqrt{\frac{m_\phi}{m_\chi}}{\vec q}'\Big)^2+m_\phi(2m_\chi-m_\phi)}. 
\eea
Then, the resulting Yukawa potential between $\chi({\bar\chi})$ and $\phi$ are attractive (repulsive).  The overall signs in the scattering amplitudes for fermion and anti-fermion differ because of Fermi statistics under the exchange of fermions.
But, for $m_\phi=2m_\chi$, the effective mediator mass vanishes for both fermion and anti-fermion, so does the overall coefficient of the scattering amplitude. In this case, we need to keep the derivative terms in the numerator of the scattering amplitude, so the scattering amplitude gets momentum-suppressed. 
Therefore,  this case is not relevant for self-resonant dark matter.

\underline{Case with pseudo-scalar dark matter}

We consider a two-component dark matter with  fermion dark matter $\chi$ and pseudo-scalar dark matter $a$, with the following Lagrangian,
\bea
{\cal L}_\chi= \frac{1}{2}(\partial_\mu a)^2-\frac{1}{2} m^2_a a^2+{\bar\chi}(i\gamma^\mu\partial_\mu-m_\chi)\chi - i \lambda_\chi a {\bar\chi} \gamma^5\chi. 
\eea
In this case, we can consider a similar elastic scattering process, $\chi a\to \chi a$ and its complex conjugate process.

For elastic scattering processes, $\chi(q)a(p)\to \chi(q')a(p')$ and ${\bar\chi}(q)a(p)\to {\bar\chi}(q')a(p')$,  the corresponding scattering amplitude at tree level is given in the non-relativistic limit, as follows,
\bea
{\tilde\Gamma}_\chi(p,q;p',q') = \frac{\mp 2\lambda^2_\chi m_\chi m_a}{\Big(\sqrt{\frac{m_\chi}{m_a}} {\vec p}-\sqrt{\frac{m_a}{m_\chi}}{\vec q}'\Big)^2+m_a(2m_\chi-m_a)}. 
\eea
In this case, there is no overall suppression of the scattering amplitude for $m_a=2m_\chi$. Similarly, the overall signs in the scattering amplitudes for fermion and anti-fermion are different.

As a result, the effective Yukawa-like potentials in the Bethe-Salpeter equation  for the $\chi-a$  and ${\bar\chi}-a$ systems are given by
\bea
V(r) = \pm (-1)^l\, \frac{\alpha_\chi}{r}\, e^{-M_\chi r}
\eea
with
\bea
\alpha_\chi &\equiv & \frac{m_a \lambda^2_\chi}{8\pi m_\chi}, \\
M_\chi &\equiv & m_a\sqrt{2-\frac{m_a}{m_\chi}}. 
\eea
Therefore, the Yukawa potential for the $\chi-a$ system is repulsive for $s$-wave with $l=0$, but it is attractive for $p$-wave with $l=1$. 
On the other hand, the Yukawa potential for the ${\bar\chi}-a$ system is attractive for $s$-wave, but it is repulsive for $p$-wave. 
As a result, the Sommerfeld enhancement and the non-perturbative self-scattering for dark matter are possible only from ${\bar\chi}-a$ scattering.

\underline{Case with vector dark matter}

We now consider a two-component dark matter with  fermion dark matter $\chi$ and vector dark matter $Z'$, with the following Lagrangian,
\bea
{\cal L}_\chi= -\frac{1}{4} Z'_{\mu\nu} Z^{'\mu\nu}+\frac{1}{2} m^2_{Z'} Z'_\mu Z^{\prime \mu} +{\bar\chi}(i\gamma^\mu\partial_\mu-m_\chi)\chi -  g_{Z'}  Z'_\mu  {\bar\chi} \gamma^\mu\chi
\eea
where $g_{Z'}$ is the gauge coupling for fermion dark matter and $Z'_{\mu\nu}=\partial_\mu Z'_\nu-\partial_\nu Z'_\mu$.
In this case, we can consider a similar elastic scattering process, $\chi Z'\to \chi Z'$ and its complex conjugate process.

The tree-level scattering amplitudes for  $\chi(q)Z'(p)\to \chi(q')Z'(p')$ and ${\bar\chi}(q)Z'(p)\to {\bar\chi}(q')Z'(p')$ are approximated in the non-relativistic limit as
\bea
{\tilde\Gamma}_\chi(p,q;p',q') = \frac{\mp 2g^2_{Z'} m_\chi m_{Z'}}{\Big(\sqrt{\frac{m_\chi}{m_{Z'}}} {\vec p}-\sqrt{\frac{m_{Z'}}{m_\chi}}{\vec q}'\Big)^2+m_{Z'}(2m_\chi-m_{Z'})}.
\eea
In this case, similarly to the case with fermion and pseudo-scalar dark matter,  there is no overall suppression of the scattering amplitude for $m_{Z'}=2m_\chi$, and the overall signs for elastic scattering amplitudes  for dark fermion and anti-fermion are opposite.

As a result, the effective Yukawa-like potentials in the Bethe-Salpeter equation  for the $\chi-Z'$  and  ${\bar\chi}-Z'$ systems are given by
\bea
V(r) = \pm (-1)^l\, \frac{\alpha_\chi}{r}\, e^{-M_\chi r}
\eea
with
\bea
\alpha_\chi &\equiv & \frac{m_{Z'} g^2_{Z'}}{8\pi m_\chi}, \\
M_\chi &\equiv & m_{Z'}\sqrt{2-\frac{m_{Z'}}{m_\chi}}. 
\eea
Therefore, the Yukawa potential for the $\chi-Z'$ system is repulsive for $s$-wave, but it is attractive for $p$-wave. 
On the other hand, the Yukawa potential for the ${\bar\chi}-Z'$ system is attractive for for $s$-wave, but it is repulsive for $p$-wave. 
As a result, there are similar Sommerfeld or non-perturbative enhancements only from ${\bar\chi}-Z'$ scattering.

\underline{Case with axial-vector dark matter}

We also consider a two-component dark matter with  fermion dark matter $\chi$ and axial-vector dark matter $A'$, with the following Lagrangian,
\bea
{\cal L}_\chi= -\frac{1}{4} A'_{\mu\nu} A^{\prime \mu\nu}+\frac{1}{2} m^2_{A'} A'_\mu A^{\prime\mu} +{\bar\chi}(i\gamma^\mu\partial_\mu-m_\chi)\chi -  g_{A'}  A'_\mu  {\bar\chi} \gamma^\mu \gamma^5 \chi. 
\eea
In this case, we can consider a similar elastic scattering process, $\chi A'\to \chi A'$ and its complex conjugate process.

In the non-relativistic limit, we can approximate the tree-level scattering amplitudes for  $\chi(q)A'(p)\to \chi(q')A'(p')$ and  ${\bar\chi}(q)A'(p)\to {\bar\chi}(q')A'(p')$  to
\bea
{\tilde\Gamma}_\chi(p,q;p',q') = \frac{\pm 2g^2_{A'} m_\chi(2m_\chi-m_{A'})}{\Big(\sqrt{\frac{m_\chi}{m_{A'}}} {\vec p}-\sqrt{\frac{m_{A'}}{m_\chi}}{\vec q}'\Big)^2+m_{A'}(2m_\chi-m_{A'})}. 
\eea
Thus, similarly to the case with fermion and scalar dark matter, the overall coefficient of the scattering amplitude vanishes when the effective mediator mass is zero for $m_{A'}=2m_\chi$.
So, this case is not relevant for self-resonant dark matter.

\subsection{Two-component vector dark matter}

We can also consider models for two-component vector dark matter, $X^\mu_i (i=1,2,3)$. In this case, the non-abelian gauge interactions such as a broken $SU(2)$ \cite{vsimp1,vsimp2} can give rise to the necessary triple coupling for dark matter resonance, 
\bea 
{\cal L}_X= -\frac{1}{4} {\vec X}_{\mu\nu}\cdot {\vec X}^{\mu\nu} - \frac{1}{2}g_X (\partial_\mu {\vec X}_\nu-\partial_\nu{\vec X}_\mu)\cdot ({\vec X}^\mu\times {\vec X}^\nu) +\cdots
\eea
where $ {\vec X}_{\mu\nu}=\partial_\mu {\vec X}_\nu-\partial_\nu {\vec X}_\mu$, $g_X$ is the gauge coupling and the ellipsis means quartic self-interactions. Then, for $m_{X_1}=m_{X_2}\equiv m_X> m_{X_3}/2$, the model provides two-component dark matter too. The case can be generalized to the case where the gauge boson masses are different.

For elastic scattering process, $X (q)X_3(p)\to X(q') X_3(p')$, with $X_\mu=\frac{1}{\sqrt{2}}(X_{1,\mu}+i X_{2,\mu})$, the squared  tree-scattering amplitude at tree level is given by
\bea
{\tilde\Gamma}_X(p,q;p',q')= \frac{-2g^2_X m_{X_3}(2m_X-m_{X_3}) (m_X+m_{X_3})/m^2_X}{\Big(\sqrt{\frac{m_\chi}{m_X}} {\vec p}-\sqrt{\frac{m_X}{m_\chi}}{\vec q}'\Big)^2+m_X(2m_X-m_{X_3})}.
\eea
The scattering amplitude for $X^\dagger X_3\to X^\dagger X_3$ is the same as for  $X X_3\to X X_3$.
Thus, similarly to the case with fermion and scalar (axial-vector) dark matter, the overall coefficient of the scattering amplitude vanishes when the effective mediator mass is zero for $m_{X_3}=2m_X$.  So, this case is not relevant for self-resonant dark matter.

\section{Thermal relics from self-resonant dark matter}

We consider the relic density for self-resonant dark matter and the constraints on the model and its extensions with an mediator particle. 

For two-component scalar dark matter with $m_2>m_1$, the $\phi_2$ dark matter can annihilate by the $2\to 2$ annihilation process, $\phi_2\phi_2\to \phi_1\phi^*_1$. 
In our model,  the same triple coupling between two dark matter components, which is responsible for the $u$-channel resonance, determines the annihilation cross section for $\phi_2\phi_2\to \phi_1\phi^*_1$ as
\bea
\langle \sigma v\rangle_{\phi_2\phi_2\to \phi_1\phi^*_1}= \frac{2g^4 m^4_1}{\pi m^6_2}\sqrt{1-\frac{m^2_1}{m^2_2}} \equiv \frac{\alpha^2_1}{m^2_1},
\eea
and the annihilation cross section for $\phi_1\phi_1\phi^*_1\to \phi_1 \phi_2$ is  given in eq.~(\ref{32tree}) without a Sommerfeld enhancement.
For $m_2\simeq 2m_1$, we get $\alpha_1\simeq 1.16\alpha$ with $\alpha=g^2/(4\pi)$.

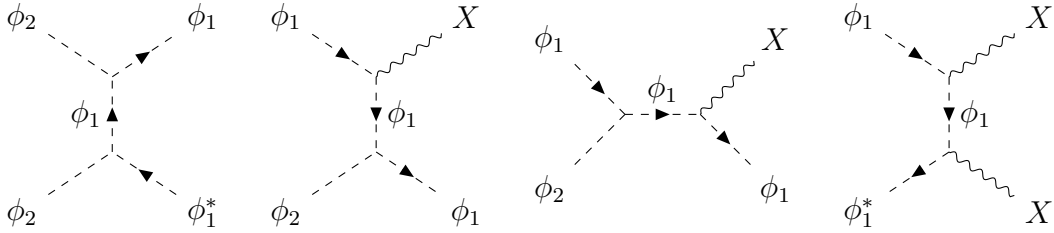
\begin{figure}[!t]
\begin{center}
\begin{equation}
\begin{tikzpicture} [baseline=($0.5*(ini1)+0.5*(ini2)$)]
	\begin{feynman} [inline=($0.5*(ini1)+0.5*(ini2)$)]
		\vertex (ini1) { \( \phi_2 \) }	;	\vertex [below = 2.6cm of ini1](ini2) { \( \phi_2 \) }	;
		\vertex [below right = 0.8cm and 1.2cm of ini1] (int1)	; \vertex [above right = 0.8cm and 1.2cm of ini2](int2) ;
		\vertex [right = 2.4cm of ini1](f1) { \( \phi_1 \) } ;	\vertex [right = 2.4cm of ini2](f2) { \( \phi^*_1 \) } ;
		
		\diagram*{
			(ini1) -- [scalar] (int1) ,
			(ini2) -- [scalar] (int2),
			(f2) -- [charged scalar] (int2) -- [charged scalar, edge label =  \( \phi_1 \)] (int1) -- [charged scalar] (f1),
		};
	\end{feynman}
\end{tikzpicture}\,\,\,\,\,\,
\begin{tikzpicture} [baseline=($0.5*(ini1)+0.5*(ini2)$)]
    \begin{feynman} [inline=($0.5*(ini1)+0.5*(ini2)$)]
        \vertex (ini1) { \( \phi_1 \) } ;   \vertex [below = 2.6cm of ini1](ini2) { \( \phi_2 \) }  ;
        \vertex [below right = 0.8cm and 1.2cm of ini1] (int1)  ; \vertex [above right = 0.8cm and 1.2cm of ini2](int2) ;
        \vertex [right = 2.4cm of ini1](f1) { \( X \) } ;   \vertex [right = 2.4cm of ini2](f2) { \( \phi_1 \) } ;
        
        \diagram*{
            (ini1) -- [charged scalar] (int1) -- [charged scalar, edge label =  \( \phi_1 \)] (int2) -- [charged scalar] (f2),
            (ini2) -- [scalar] (int2),
            (int1) -- [photon] (f1),
        };
    \end{feynman}
\end{tikzpicture}\,\,\,\,\,\,
\begin{tikzpicture} [baseline=($0.5*(ini1)+0.5*(ini2)$)]
	\begin{feynman} [inline=($0.5*(ini1)+0.5*(ini2)$)]
		\vertex (ini1) { \( \phi_1 \) }	;	\vertex [below = 2cm of ini1](ini2) { \( \phi_2 \) }	;
		\vertex [below right = 1cm and 1cm of ini1] (int1)	; 
		\vertex [right = of int1](int2) ;
		\vertex [right = 3cm of ini1](f1) { \( X \) } ;	\vertex [right = 3cm of ini2](f2) { \( \phi_1 \) } ;
		
		\diagram*{
			(ini1) -- [charged scalar] (int1) -- [charged scalar, edge label =  \( \phi_1 \)] (int2) -- [charged scalar] (f2),
			(ini2) -- [scalar] (int1),
			(int2) -- [photon] (f1),
		};
	\end{feynman}
\end{tikzpicture}\,\,\,\,\,\,
\begin{tikzpicture} [baseline=($0.5*(ini1)+0.5*(ini2)$)]
	\begin{feynman} [inline=($0.5*(ini1)+0.5*(ini2)$)]
		\vertex (ini1) { \( \phi_1 \) }	;	\vertex [below = 2.6cm of ini1](ini2) { \( \phi^*_1 \) }	;
		\vertex [below right = 0.8cm and 1.2cm of ini1] (int1)	; \vertex [above right = 0.8cm and 1.2cm of ini2](int2) ;
		\vertex [right = 2.4cm of ini1](f1) { \( X \) } ;	\vertex [right = 2.4cm of ini2](f2) { \( X \) } ;
		
		\diagram*{
			(ini1) -- [charged scalar] (int1) -- [charged scalar, edge label =  \( \phi_1 \)] (int2) -- [charged scalar] (ini2),
			(int1) -- [photon] (f1) ,
			(int2) -- [photon] (f2),
		};
	\end{feynman}
\end{tikzpicture} \nonumber
\end{equation}
\end{center}
\caption{Feynman diagrams for the annihilation of self-resonant dark matter. Here, $X$ can be a dark photon or a dark Higgs.}
\label{22ann}
\end{figure}

In order to reduce the relic abundance for $\phi_1$ sufficiently during freeze-out, we need to introduce an additional annihilation channel for $\phi_1$ such as $\phi_1\phi^*_1\to XX$  with an extra mediator $X$ having mass $m_X<m_1$. For instance, we can take the extra mediator $X$ to be a dark photon $Z'$ or a dark Higgs $h'$.  Then, there also appears a semi-annihilation channel for $\phi_2$ through $\phi_1\phi_2\to \phi_1 X$ and its conjugate, for $m_X<m_2$. We note that $\phi_2$ is still stable, as far as $m_2<2m_1+m_X$ for on-shell $X$ and $m_2<2m_1+2m_f$ for off-shell $X$ decaying into a pair of the Standard Model (SM) particles, $f{\bar f}$.  
The Sommerfeld enhancement with the $u$-channel resonance is relevant only for  the semi-annihilation processes, $\phi_1\phi_2\to \phi_1 X$. In Fig.~\ref{22ann}, assuming a mediator coupling only to $\phi_1$ as well as  the triple coupling between two dark matter components, we show the possible Feynman diagrams relevant for the $2\to 2$ annihilation of self-resonant dark matter.

Furthermore, in the presence of the mediator $X$, dark matter can also annihilate through $3\to 2$ channels, $\phi_1\phi_1\phi^*_1\to \phi_1 \phi_2$ and $\phi_1\phi_1\phi^*_1\to \phi_1 X$. For light dark matter below sub-GeV scale, those higher order annihilation channels become relevant \cite{simp,simp2,axionportal,simpz3,simprelic,review,vsimp1,vsimp2}. In this case, such a sub-GeV dark matter can have a large self-interaction at the perturbative level. 
However, in the case of $u$-channel resonances, those $3\to 2$ channels are subdominant for determining the relic density, as will be discussed shortly.

\subsection{Dark photon portals}

In the case that $X$ is a massive dark gauge boson with mass $m_X$ and interaction term with $\phi_1$, ${\cal L}_{\rm int}=-ig_X X_\mu (\phi_1\partial^\mu\phi^*_1-\phi^*_1\partial^\mu\phi_1)+g^2_X |\phi|^2 X_\mu X^\mu$, we obtain the corresponding annihilation cross sections for $\phi_1\phi^*_1\to XX$ and $\phi_1\phi_2\to \phi_1 X$ \cite{z3dm}, respectively, in the non-relativistic limit of the initial states, as follows,
\bea
\langle \sigma v\rangle_{\phi_1\phi^*_1\to XX} &=&\frac{g^4_X}{16\pi m^2_1}\frac{8m^4_1-8m^2_1 m^2_X+3m^4_X}{(2m^2_1-m^2_X)^2}\sqrt{1-\frac{m^2_X}{m^2_1}}\equiv \frac{2\alpha^2_2}{m^2_1},  \\
\langle \sigma v\rangle_{\phi_1\phi_2\to \phi_1 X} &=& \frac{ (2m_1+m_2) |{\cal M}_X|^2 }{32\pi m_1(m_1+m_2)^2}\sqrt{1-\frac{m^2_X}{m^2_2}}\sqrt{1-\frac{m^2_X}{(2m_1+m_2)^2}}
\nonumber \\
&&\quad\times\sqrt{1-\frac{(m_1+m_X)^2}{(m_1+m_2)^2}}\,\sqrt{1-\frac{(m_1-m_X)^2}{(m_1+m_2)^2}} \nonumber \\
&\equiv& \frac{2\alpha^2_3}{m^2_1}
\eea
where 
\bea
|{\cal M}_X|^2= \frac{4g^2 g^2_X m_1^2 m_X^2 }{m_2^2(m_2+2m_1)^2}\, \frac{(m_2^2-m_X^2)\big((m_2+2m_1)^2-m_X^2\big)}{\big(m_1(m_2+2m_1)-m_X^2\big)^2}.
\eea
Then, we find that $\phi_1\phi_2\to \phi_1 X$ channel is dominated by the $s$-wave.
The annihilation cross section for $\phi_1\phi_2\to \phi_1 X$ could be enhanced by $u$-channel Sommerfeld effects and would be strongly constrained by indirect detection for dark matter and CMB bounds, if the dark photon $X$ decays into visible particles. 

Moreover, the annihilation cross section for $\phi_1\phi_1\phi^*_1\to \phi_1 X$ is given by
\bea
(\sigma v^2)_{\phi_1\phi_1\phi^*_1\to \phi_1 X}  =\frac{|{\cal M}_{\phi_1\phi_1\phi^*_1\to \phi_1 X}|^2}{72\pi m^3_1}\sqrt{\bigg(1-\frac{m^2_X}{4m^2_1}\bigg)\bigg(1-\frac{m^2_X}{16m^2_1}\bigg)}
\eea
with
\bea
|{\cal M}_{\phi_1\phi_1\phi^*_1\to \phi_1 X}|^2&=&\frac{g^2_X m^2_X(16m^2_1-m^2_X)}{6(m^2_2-4m^2_1)^2(4m^2_1-m^2_X)(2m^2_1+m^2_X)^2}\times \nonumber \\
&&\times \bigg( 3 g^2_X (m^2_2-4m^2_1)+g^2 (2m^2_1+m^2_X)\bigg)^2.
\eea

We remark that the $\phi_1\phi_1\phi^*_1\to \phi_1 X$ channel is subdominant for WIMP-like dark matter, but it can be crucial for sub-GeV dark matter. Comments on that are in order.
For $g=0$, both $\phi_1\phi^*_1\to XX$ and the $3\to 2$ semi-annihilation cross sections with dark photon are nonzero due to the dark photon gauge coupling $g_X$. So, for sub-GeV dark matter,  we can make the $2\to 2$ annihilation Boltzmann-suppressed for $m_X\gtrsim m_1$ and determine the relic density by $\phi_1\phi^*_1\to XX$ and the $3\to 2$ semi-annihilation \cite{vsimp1}.
For $g\neq 0$ and $m_2\sim 2m_1$, which are needed for the $u$-channel resonance, we can take $m_X\gtrsim m_2$, but not only the $2\to 2$ channels but also the $\phi_1\phi_1\phi^*_1\to \phi_1 X$ channel are kinematically forbidden. So, in this case, the $2\to 2$ annihilation channels with kinematic suppression or small couplings \cite{vsimp1,vsmip2} are most important for light dark matter.

\subsection{Dark Higgs portals}

We also note that if $X=h'$ is a dark Higgs boson, with mass $m_{h'}$ and interaction term with $\phi_1$, ${\cal L}_{\rm int}=- \lambda_m |\phi'|^2 |\phi_1|^2-\lambda_\phi |\phi'|^4$ where $\phi'=(v_\phi+h')/\sqrt{2}$ with $v_\phi$ being the vacuum expectation value of the dark Higgs, then the corresponding annihilation cross sections for $\phi_1\phi^*_1\to h'h'$ and $\phi_1\phi_2\to \phi_1 h'$ \cite{z3dm}, respectively, become in the non-relativistic limit of the initial states, as follows,
\bea
\langle \sigma v\rangle_{\phi_1\phi^*_1\to h'h'} &=& \frac{\lambda^2_m}{64\pi m^2_1} \bigg(1-\frac{2\lambda_m v^2_\phi}{2m^2_1-m^2_{h'}} +\frac{6\lambda_\phi v^2_\phi }{4m^2_1-m^2_{h'}}\bigg)^2 \sqrt{1-\frac{m^2_{h'}}{m^2_1}}\equiv  \frac{2\alpha^{\prime 2}_2}{m^2_1}, \\
\langle \sigma v\rangle_{\phi_1\phi_2\to \phi_1 h'} &=&\frac{ (2m_1+m_2) |{\cal M}_{h'}|^2 }{32\pi m_1(m_1+m_2)^2}\sqrt{1-\frac{m_{h'}^2}{m^2_2}}\sqrt{1-\frac{m_{h'}^2}{(2m_1+m_2)^2}}
\nonumber \\
&&\quad\times\sqrt{1-\frac{(m_1+m_{h'})^2}{(m_1+m_2)^2}}\,\sqrt{1-\frac{(m_1-m_{h'})^2}{(m_1+m_2)^2}} \nonumber \\
&\equiv& \frac{2\alpha^{\prime 2}_3}{m^2_1}
\eea
where
\bea
 |{\cal M}_{h'}|^2=\frac{4g^2 \lambda_m^2 m_1^2 v^2_\phi}{m^2_2 (m_2+2m_1)^2}\, \frac{(m_2^2+2m_1 m_2+ m^2_{h'})^2}{\big(m_1(m_2+2m_1)-m^2_{h'}\big)^2}.
\eea
Here, we note that for the semi-annihilation,  $\phi_1\phi_2\to \phi_1 h'$, both the $u$-channel resonance due to the off-shell dark matter and the $s$-channel resonance with dark Higgs boson appear.
Since the $\phi_1\phi_2\to \phi_1 h'$ channel is dominated by the $s$-wave, it could be enhanced by $u$-channel Sommerfeld effects and would be strongly constrained by indirect detection for dark matter and CMB bounds, if the dark Higgs $h$ decays into visible particles. 

Moreover, the annihilation cross section for $\phi_1\phi_1\phi^*_1\to \phi_1 h'$ is given by
\bea
(\sigma v^2)_{\phi_1\phi_1\phi^*_1\to \phi_1 h'}  =\frac{|{\cal M}_{\phi_1\phi_1\phi^*_1\to \phi_1 h'}|^2}{72\pi m^3_1}\sqrt{\bigg(1-\frac{m^2_{h'}}{4m^2_1}\bigg)\bigg(1-\frac{m^2_{h'}}{16m^2_1}\bigg)}
\eea
with
\bea
&&|{\cal M}_{\phi_1\phi_1\phi^*_1\to \phi_1 h'}|^2=\frac{\lambda^2_m v^2_\phi}{32m^4_1(m_2^2-4m^2_1)^2(3m^2_2+4m^2_1-m^2_{h'})^2} \frac{1}{(2m^2_1+m^2_{h'})^2(4m^2_1-m^2_{h'})^4} \times \nonumber \\
&&\quad\times \bigg(4 \lambda_m m_1^2 (m_2^2 - 4 m_1^2) (3 m_2^2 + 4 m_1^2 - 
      m_{h'}^2) (32 m_1^4 - 28 m_1^2 m_{h'}^2 + 5 m_{h'}^4)  \nonumber \\
     &&\qquad+ 
   4 g^2 m_1^2 (8 m_1^4 + 2 m_1^2 m_{h'}^2 - m_{h'}^4) \Big(256 m_1^4 + 
      4 m_1^2 m_{h'}^2 + m_{h'}^4 - 3 m_2^2 (32 m_1^2 + m_{h'}^2)\Big)  \nonumber \\
      &&\qquad- 
   \lambda_m^2 (m_2^2 - 4 m_1^2) (3 m_2^2 + 4 m_1^2 - 
      m_{h'}^2) (128 m_1^4 - 46 m_1^2 m_{h'}^2 - m_{h'}^4) v_\phi^2\bigg)^2.
\eea

We comment on the $\phi_1\phi_1\phi^*_1\to \phi_1 h'$ for sub-GeV dark matter. 
For $g=0$, both $\phi_1\phi^*_1\to h'h'$ and the $3\to 2$ semi-annihilation cross sections with dark Higgs are nonzero, due to the dark Higgs-portal coupling, $\lambda_m$. So,  for the freeze-out of a sub-GeV dark matter, we can make the $2\to 2$ annihilation Boltzmann-suppressed for $m_{h'}\gtrsim m_1$ and determine the relic density by  $\phi_1\phi^*_1\to XX$ and the $3\to 2$ semi-annihilation \cite{vsimp1}, similarly as for dark photon portals. But, For $g\neq 0$ and $m_2\sim 2m_1$, which are needed for the $u$-channel resonance, we can take $m_X\gtrsim m_2$, but not only the $2\to 2$ channels but also the $\phi_1\phi_1\phi^*_1\to \phi_1 h'$ channel are kinematically forbidden. So, in this case, the $2\to 2$ annihilation channels with kinematic suppression or small couplings \cite{vsimp1,vsimp2} are most important for light dark matter.

\subsection{Boltzmann equations and dark matter relic abundances}

We denote the number densities for $\phi_1, \phi^*_1$ and $\phi_2$ by $n_{\phi_1},  n_{\phi^*_1}$ and $n_{\phi_2}$, in order, and assume $n_{\phi_1}  =n_{\phi^*_1}$ for CP conservation in the dark sector. We also assume that $X$ is in thermal equilibrium with the SM bath, but it couples to the SM weakly. So, we don't consider the direct annihilation of dark matter into the SM particles. 

In the presence of the triple coupling between dark matter particles and the mediator coupling, we consider the relevant annihilation channels for dark matter. Then, the number densities for self-resonant dark matter are governed by the following Boltzmann equations,
\bea
{\dot n}_{\phi_1} + 3H n_{\phi_1} &=& \langle \sigma v\rangle_{\phi_2\phi_2\to \phi_1\phi^*_1} \Big(n^2_{\phi_2}- n^2_{\phi_1}\Big) \nonumber \\
&&- \langle\sigma v\rangle_{\phi_1\phi^*_1\to XX} \Big(n^2_{\phi_1} -(n^{\rm eq}_{\phi_1})^2 \Big) \nonumber \\
&&-2\langle \sigma v^2 \rangle_{\phi_1\phi_1\phi^*_1\to \phi_1 \phi_2} n_{\phi_1} \bigg(n^2_{\phi_1}-\frac{(n^{\rm eq}_{\phi_1})^2}{n^{\rm eq}_{\phi_2}}\, n_{\phi_2}  \bigg) \nonumber \\
&&-2\langle \sigma v^2 \rangle_{\phi_1\phi_1\phi^*_1\to \phi_1 X} n_{\phi_1} (n^2_{\phi_1} - (n^{\rm eq}_{\phi_1})^2), \\
 {\dot n}_{\phi_2} + 3H n_{\phi_2} &=& -2\langle \sigma v\rangle_{\phi_2\phi_2\to \phi_1\phi^*_1} \Big(n^2_{\phi_2} -n^2_{\phi_1} \Big) \nonumber \\
 &&-\langle\sigma v\rangle_{\phi_1\phi_2\to \phi_1 X} \, n_{\phi_1}\Big(n_{\phi_2} -n^{\rm eq}_{\phi_2}\Big) \nonumber \\
&&+\langle \sigma v^2 \rangle_{\phi_1\phi_1\phi^*_1\to \phi_1 \phi_2} n_{\phi_1} \bigg(n^2_{\phi_1}-\frac{(n^{\rm eq}_{\phi_1})^2}{n^{\rm eq}_{\phi_2}}\, n_{\phi_2}  \bigg)
\eea
where $n^{\rm eq}_{\phi_1}, n^{\rm eq}_{\phi_2}$ are the number densities at equilibrium and we assumed that $X$ is in thermal equilibrium with the SM plasma.
Then, from $n_1\equiv n_{\phi_1}+n_{\phi^*_1}=2n_{\phi_1}$ and $n_2\equiv n_{\phi_2} $, we can rewrite the above Boltzmann equations in the following form,
\bea
{\dot n}_1 + 3H n_1 &=&\frac{1}{2} \langle \sigma v\rangle_{\phi_2\phi_2\to \phi_1\phi^*_1} \Big(4n^2_2- n^2_1\Big) \nonumber \\
&&-\frac{1}{2} \langle\sigma v\rangle_{\phi_1\phi^*_1\to XX} \Big(n^2_1 -(n^{\rm eq}_1)^2 \Big) \nonumber \\
&&-\frac{1}{2}\langle \sigma v^2 \rangle_{\phi_1\phi_1\phi^*_1\to \phi_1 \phi_2} n_1 \bigg(n^2_1-\frac{(n^{\rm eq}_1)^2}{n^{\rm eq}_2}\, n_2  \bigg) \nonumber \\
&&-\frac{1}{2}\langle \sigma v^2 \rangle_{\phi_1\phi_1\phi^*_1\to \phi_1 X} n_1(n^2_1 - (n^{\rm eq}_1)^2), \\
 {\dot n}_2 + 3H n_2 &=& -\frac{1}{2}\langle \sigma v\rangle_{\phi_2\phi_2\to \phi_1\phi^*_1} \Big(4n^2_2-n^2_1 \Big) \nonumber \\
 &&- \frac{1}{2}\langle\sigma v\rangle_{\phi_1\phi_2\to \phi_1 X} \, n_1\Big(n_2 -n^{\rm eq}_2\Big) \nonumber \\
 &&+\frac{1}{8}\langle \sigma v^2 \rangle_{\phi_1\phi_1\phi^*_1\to \phi_1 \phi_2} n_1 \bigg(n^2_1-\frac{(n^{\rm eq}_1)^2}{n^{\rm eq}_2}\, n_2 \bigg).
\eea

 \begin{figure}[tbp]
  \begin{center}
    \includegraphics[height=0.45\textwidth]{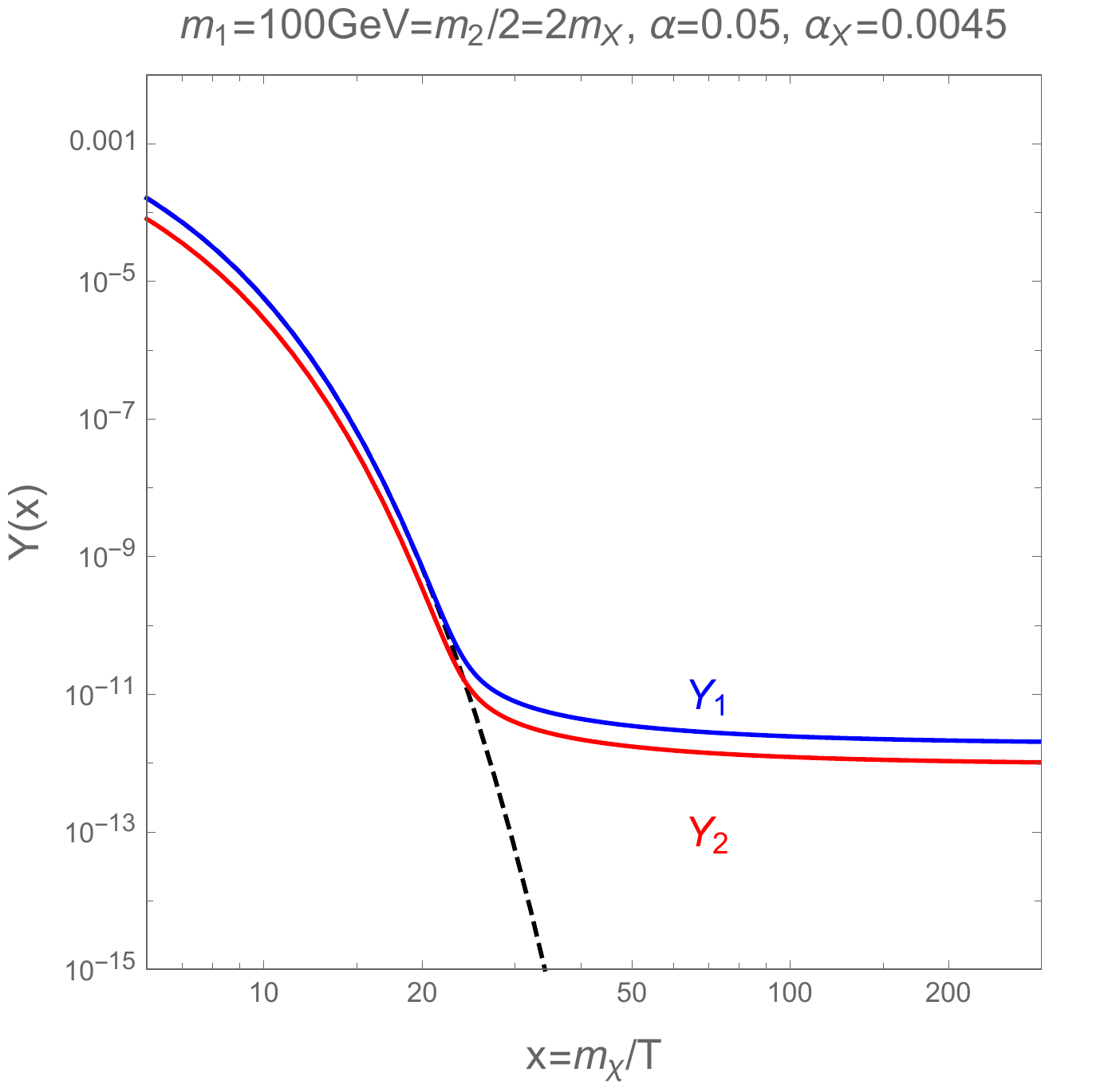}\,\,\,
        \includegraphics[height=0.45\textwidth]{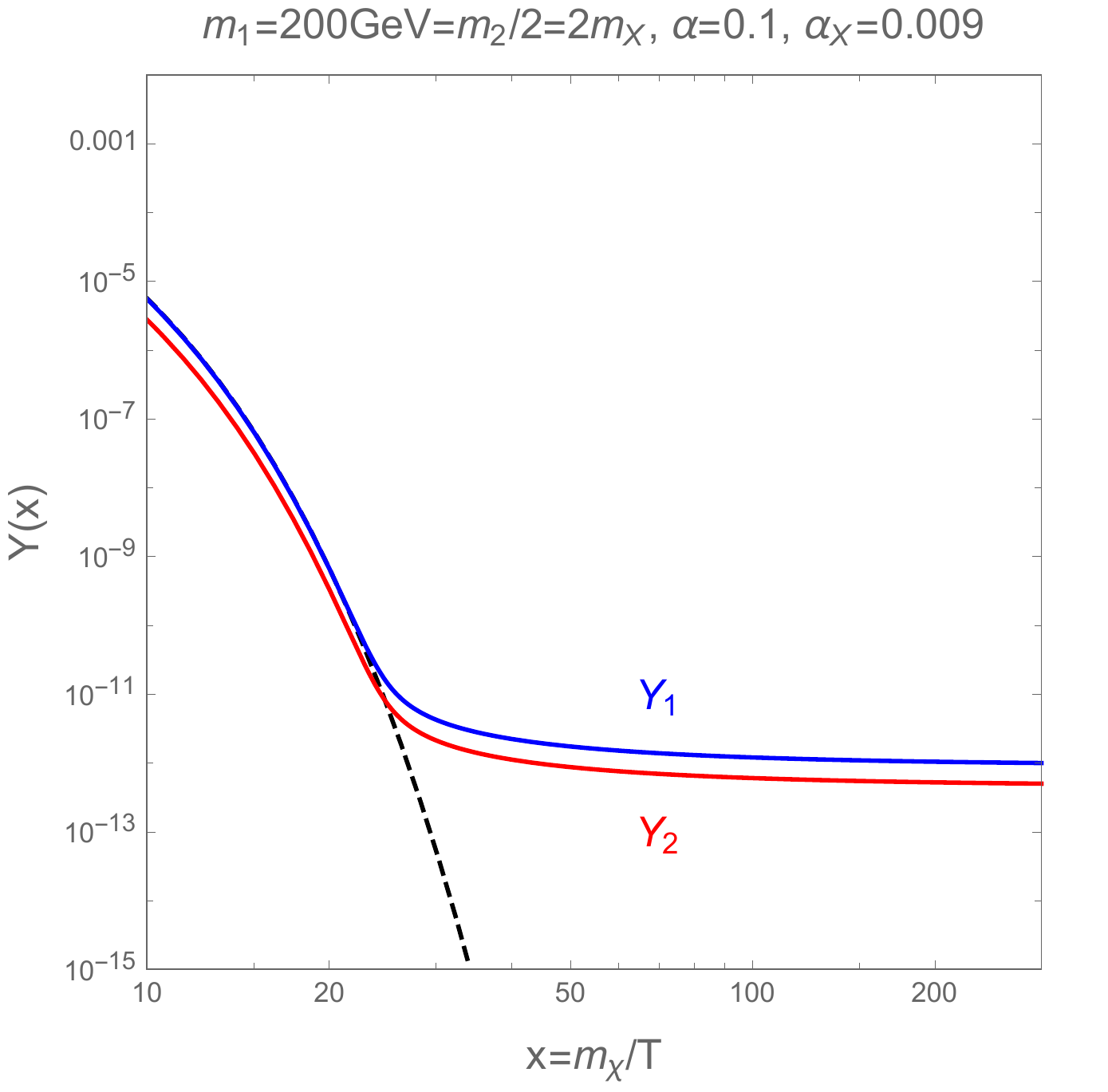}
  \end{center}
  \caption{The relic abundances for two-component dark matter for $\phi_1$ and $\phi_2$, assisted with dark gauge boson $X$, shown in blue and red lines, respectively.  We  took  $m_1=m_2/2=100\,{\rm GeV}=2m_X$ and $\alpha=g^2/(4\pi)=0.05, \alpha_X=g^2_X/(4\pi)=0.0045$, on left (B1), and  $m_1=m_2/2=200\,{\rm GeV}=2m_X$ and $\alpha=0.1, \alpha_X=0.009$, on right (B2). The dashed line corresponds to the case for the equilibrium of $\phi_1$. }
  \label{relic}
\end{figure}

Consequently, in terms of the abundances, $Y_i\equiv n_i/s$, with $s$ being the entropy density, and making a change of variables from time to $x=m_1/T$, and parametrizing the averaged annihilation cross sections by $ \langle \sigma v\rangle_{\phi_2\phi_2\to \phi_1\phi^*_1}=2\sigma_1(x)$, $ \langle\sigma v\rangle_{\phi_1\phi^*_1\to XX} =2\sigma_2(x)$,  $\langle\sigma v\rangle_{\phi_1\phi_2\to \phi_1 X} =2\sigma_3(x)$,  $\langle\sigma v^2\rangle_{\phi_1\phi_1\phi^*_1\to \phi_1 \phi_2} =2\beta_1(x)$,  and $\langle\sigma v^2\rangle_{\phi_1\phi_1\phi^*_1\to \phi_1 X} =2\beta_2(x)$, we can recast the above Boltzmann equations \cite{review} into
\bea
\frac{dY_1}{dx} &=&\lambda_1\, x^{-2} \Big(4 Y^2_2- Y^2_1 \Big)  -\lambda_2\, x^{-2} \Big( Y^2_1- (Y^{\rm eq}_1)^2\Big) \nonumber \\
&&-\rho_1 x^{-5} Y_1 \bigg( Y^2_1- \frac{(Y^{\rm eq}_1)^2}{Y^{\rm eq}_2}\, Y_2 \bigg)-\rho_2 x^{-5} Y_1 \Big( Y^2_1- (Y^{\rm eq}_1)^2 \Big), \\ 
\frac{dY_2}{dx} &=&-\lambda_1\, x^{-2} \Big(4 Y^2_2- Y^2_1 \Big)  -\lambda_3\,  x^{-2} Y_1 \Big( Y_2- Y^{\rm eq}_2\Big) \nonumber \\
&&+\frac{1}{4} \rho_1 x^{-5}  Y_1 \bigg( Y^2_1- \frac{(Y^{\rm eq}_1)^2}{Y^{\rm eq}_2}\, Y_2 \bigg)
\eea
where 
\bea
\lambda_i &=& \frac{s(m_1)}{H(m_1)}\,\sigma_i  = 1.329 (g_{*s}/g^{1/2}_*) M_P\, m_1 \sigma_i, \\
\rho_i &=&\frac{s^2(m_1)}{H(m_1)}\,\beta_i =0.583 (g^2_{*s}/g^{1/2}_*) M_P\, m^4_1\,\beta_i. 
\eea
Here, we note that  the Sommerfeld factor is included in the full cross sections  and the abundances at equilibrium are given by
\bea
Y^{\rm eq}_1(x) =  \frac{45 x^2}{2 g_{*s} \pi^4}\, K_2(x), \quad Y^{\rm eq}_2(x)= \frac{45 m^2_2 x^2}{4 g_{*s} \pi^4 m^2_1}\, K_2\Big(\frac{m_2 x}{m_1}\Big).
\eea

The relic abundance for two-component dark matter \cite{review} is given by
\bea
\Omega_{\rm DM}h^2 = \Omega_1 h^2 +\Omega_2 h^2  
\eea
with
\bea
\Omega_i h^2= 0.2745 \bigg(\frac{Y_{i,\infty}}{10^{-11}} \bigg)\bigg(\frac{m_i}{100\,{\rm GeV}}\bigg)
\eea
where $Y_{i,\infty}$ are the abundances at present, typically determined by $T=T_{i,F}$ with $T_{i,F}=m_i/x_i$ being the freeze-out temperatures for $\phi_1$ and $\phi_2$.

For WIMP-like dark matter, the $2\to 2$ (semi-)annihilation channels are dominant for determining the relic density. For $g\gg g_X$, there is a hierarchy of the annihilation cross sections, $\langle\sigma v\rangle_{\phi_2\phi^*_2\to \phi_1\phi^*_1}\gg \langle\sigma v\rangle_{\phi_1\phi^*_1\to XX},  \langle\sigma v\rangle_{\phi_1\phi_2\to \phi_1 X}$. Thus, in this case, we find that the equilibrium condition between two dark matter scalars is maintained by $\phi_2\phi^*_2\to \phi_1\phi^*_1$ until $\phi_1\phi^*_1\to XX$ and $\phi_1\phi_2\to \phi_1 X$ processes are frozen, thus giving rise to $Y_1\simeq 2Y_2$ at freeze-out. Therefore, two dark matter scalars contribute equally to the total dark matter relic density.

In Fig.~\ref{relic}, we depict the relic abundances for two-component WIMP-like dark matter, $\phi_1$ and $\phi_2$, as a function of $x=m_1/T$, in blue and red lines, respectively. We take two benchmarks with $g\gg g_X$, for instance, $m_1=m_2/2=100\,{\rm GeV}=2m_X$ and $\alpha=g^2/(4\pi)=0.05, \alpha_X=g^2_X/(4\pi)=0.0045$ on left (B1), and $m_1=m_2/2=200\,{\rm GeV}=2m_X$ and $\alpha=0.1, \alpha_X=0.009$ on right (B2). We find that a sizable triple interaction between $\phi_1$ and $\phi_2$ shows that the dark matter densities for $\phi_1$ and $\phi_2$ are equally abundant. In both cases in Fig.~\ref{relic}, the total relic density for $\phi_1$ and $\phi_2$ accounts for the observed density for dark matter.

 \begin{figure}[tbp]
  \begin{center}
    \includegraphics[height=0.45\textwidth]{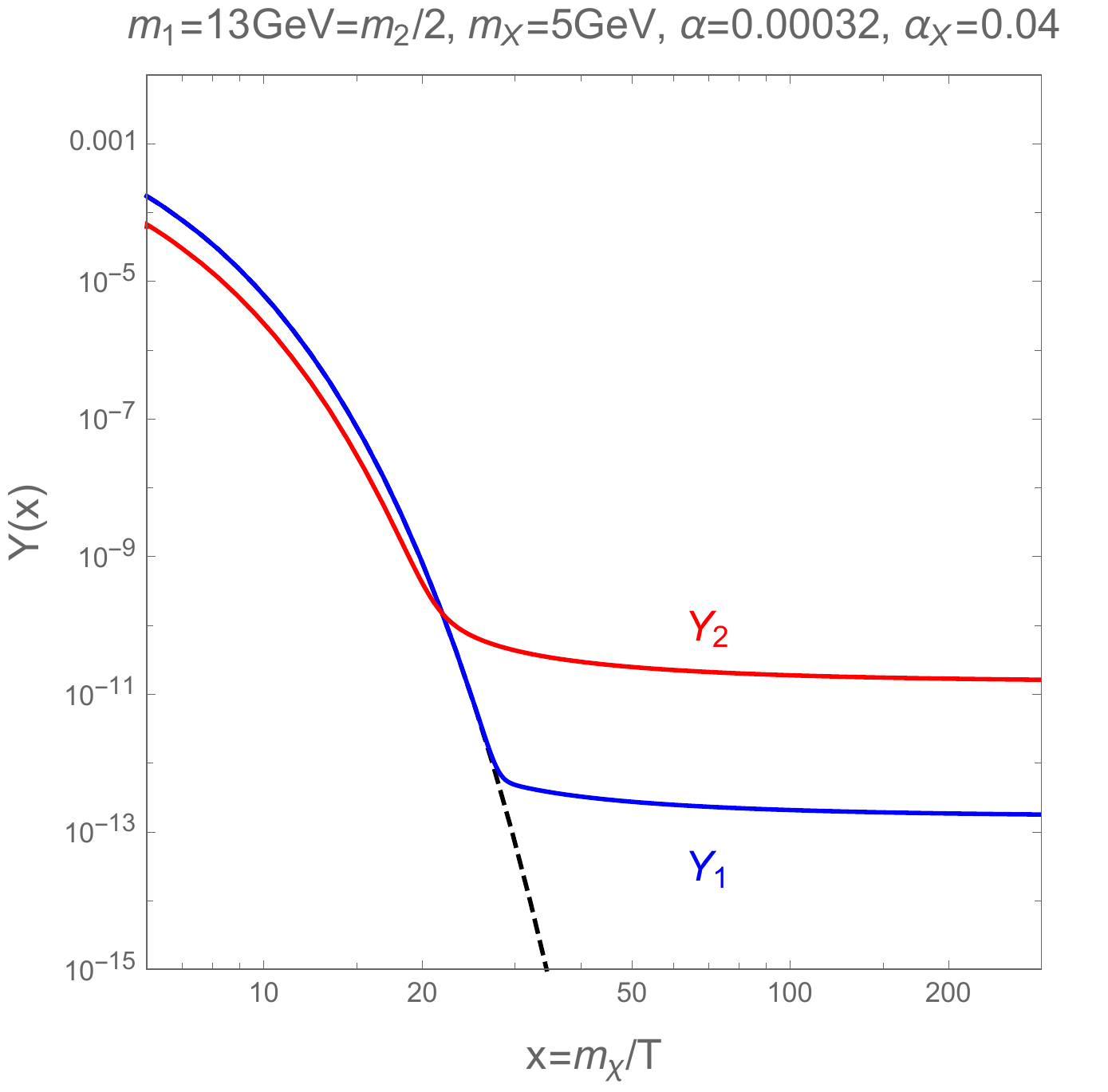}\,\,\,
        \includegraphics[height=0.45\textwidth]{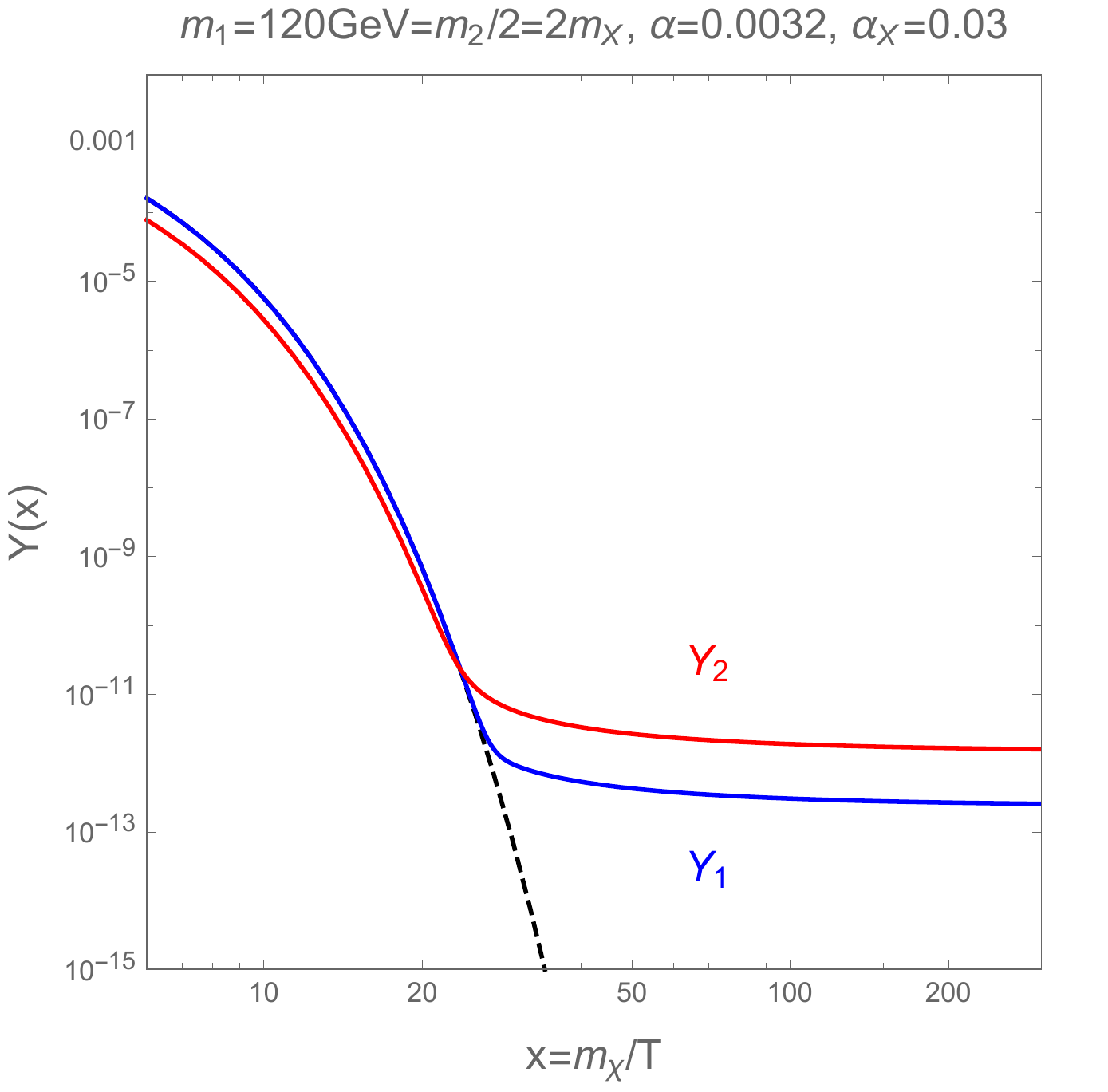}
  \end{center}
  \caption{The same as in Fig.~\ref{relic}. We  took  $m_1=m_2/2=13\,{\rm GeV}, m_X=5\,{\rm GeV}$ and $\alpha=0.00032, \alpha_X=0.04$, on left (B3), and  $m_1=m_2/2=120\,{\rm GeV}=2m_X$ and $\alpha=0.0032, \alpha_X=0.03$, on right (B4).  }
  \label{relic2}
\end{figure}

On the other hand, we also remark the case for $g\ll g_X$. In this case, the dark matter abundance for $\phi_2$  is  determined to be close to the one in thermal equilibrium at freeze-out temperature while $\phi_1$ is still in thermal equilibrium. Then, the dark matter abundance for $\phi_1$ is consequently determined by $\phi_1\phi^*_1\to XX$ and it becomes a sub-dominant component of dark matter. 

In Fig.~\ref{relic2}, we also show the relic abundances for two-component WIMP-like dark matter, $\phi_1$ and $\phi_2$, similarly to Fig.~\ref{relic}. But, now we take two benchmarks with $g\ll g_X$,  and choose $m_1=m_2/2=13\,{\rm GeV}, m_X=5\,{\rm GeV}$ and $\alpha=0.00032, \alpha_X=0.04$ on left (B3), and  $m_1=m_2/2=120\,{\rm GeV}=2m_X$ and $\alpha=0.0032, \alpha_X=0.03$ on right (B4). In this case, the dark matter abundance for $\phi_1$ is suppressed due to the delayed freeze-out, so it occupies $0.5\%$ and $7.5\%$ of the observed relic density, for the left and right plots in Fig.~\ref{relic2}, respectively.

\subsection{Indirect detection for self-resonant dark matter}

As the semi-annihilation of dark matter, $\phi_1\phi_2\to \phi_1 X$, and its complex conjugate, can be enhanced by the $u$-channel Sommerfeld effects, it can be important to consider the bounds from indirect detection experiments for dark matter, such as Fermi-LAT, AMS-02, HESS, CMB, etc. But, the indirect detection bounds depend on the decay channels of the mediator $X$. 
In the following, we discuss the case for the $s$-wave annihilation for  $\phi_1\phi_2\to \phi_1 X$ with $X$ decaying into a pair of leptons, $f{\bar f}$.

\underline{Indirect detection from semi-annihilation}

Suppose that dark matter annihilation is enhanced by $\phi_1\phi_2\to \phi_1 X$ with $X$ decaying into a pair of the SM leptons. Then, we consider the energy spectrum of the leptons. 
Ignoring the lepton masses, we get the energy of the leptons in the rest frame of the $X$ particle as ${\bar E}_f=\frac{m_X}{2}$. But, in the galactic center of mass frame, the $X$ particle has a nonzero velocity $v_X$, given by
\bea
v_X= \frac{p_X}{E_X}
\eea
with
\bea
p_X&=& \frac{m_2(2m_1+m_2)}{2(m_1+m_2)}\,\bigg(1-\frac{m^2_X}{m^2_2}\bigg)^{1/2} \Big(1-\frac{m^2_X}{(2m_1+m_2)^2} \Big)^{1/2},  \label{Xmom} \\
E_X &=& \frac{m^2_2+2m_1 m_2 +m^2_X}{2(m_1+m_2)}.
\eea
Then, the energy of the leptons in the galactic center of mass frame is boosted \cite{indirect} to
\bea
E_f=\frac{1}{\gamma_X}\, {\bar E}_f \,(1-v_X\,\cos\theta)^{-1}
\eea
where $\theta$ is the angle between the velocity of the $X$ particle and the lepton velocity, and 
\bea
\gamma_X=\frac{1}{\sqrt{1-v^2_ X}}=\frac{E_X}{m_X}. 
\eea
Therefore, the leptons can carry the energies between $E^-_f$ and $E^+_f$ with
\bea
E^\mp_f= \sqrt{\frac{1\mp v_X}{1\pm v_X}}\, \frac{m_X}{2}, \label{leptonE}
\eea
and  the energy spectrum for the leptons is box-shaped \cite{indirect} with the width being given by $\Delta E_f= E^+_f- E^-_f= v_X \gamma_X\, m_X$. 

As a consequence, we obtain the differential flux for the leptons coming from $\phi_1\phi_2\to \phi_1 X$ and its complex conjugate as
\bea
\frac{d\Phi_f}{dE_f} = \frac{1}{8\pi m_1 m_2}\, \langle\sigma v\rangle_{\phi_1\phi_2\to \phi_1 X} \, \frac{dN_f}{d E_f} \, r_1(1-r_1)\, J
\eea
where the differential number of leptons is given by
\bea
\frac{d N_f}{dE_f} = \frac{1}{E^+_f -E^-_f}\,\Theta(E_f-E^-_f) \Theta(E^+_f -E_f)\, {\rm BR}(X\to f{\bar f}),
\eea
$r_1=\Omega_1/\Omega_{\rm DM}$, and the $J$-factor is 
\bea
J=\frac{1}{\Delta \Omega} \int_{\Delta\Omega} d\Omega \int_{\rm l.o.s.} ds\,  \rho^2_{\rm DM}.
\eea
Then, positrons coming from the DM annihilation are strongly constrained by the AMS-02 experiment up to the energies of $m_2/2\simeq 10\,{\rm GeV}$ \cite{AMS02}. 

There are also strong constraints on the case with $X\to e^+ e^-$ from diffuse gamma-ray searches in dwarf spheroidal galaxies \cite{dwarf}. 
Moreover, if the mediator $X$ decays into a pair of photons, the resulting photons coming from  $\phi_1\phi_2\to \phi_1 X$ are box-shaped \cite{indirect} with the same energy interval  between $E^-_f$ and $E^+_f$ in eq.~(\ref{leptonE}). 
In this case, the line-like gamma-ray searches from Fermi-LAT, HESS, etc \cite{gammaline}, are relevant for constraining the $u$-channel Sommerfeld effects.

\underline{Effects at CMB recombination}

The dark matter annihilation, $\phi_1\phi_2\to \phi_1 X$, can be also constrained by the CMB bound \cite{planck15}. We define the effective efficiency parameter for CMB recombination \cite{slatyer3} as
\bea
f_{\rm eff}(m_2)= \frac{\int^{m_2/2}_0 dE_e \, E_e\, 2f^{e^+e^-}_{\rm eff}\, \frac{dN_e}{dE_e}}{m_2}
\eea
where $f^{e^+e^-}_{\rm eff}$ is the efficiency factor\footnote{Similar efficiency factors can be also introduced for other leptons or photons \cite{slatyer3}.} for $X\to e^+e^-$, and we took the range of the lepton energy to $m_2/2$ when two leptons coming from the $X$ decay are back-to-back.

The typical DM velocity during recombination is $\langle v^2_2\rangle=\frac{m_1^2}{(m_1+m_2)^2}\,\langle v^2\rangle=3T^2_{\rm rec}/(m_2 T_{\rm kd})$ with $T_{\rm rec}=T_{\rm kd}(z_{\rm rec}/z_{\rm kd})$. For the usual WIMP dark matter, we have $10\,{\rm MeV}\lesssim T_{\rm kd}\lesssim 1\,{\rm GeV}$. But, in our case, dark matter freeze-out depends on dark sector interactions as discussed in the previous section, so the kinetic decoupling is independent of the freeze-out condition. Since  the kinetic decoupling temperature is determined by the couplings of the $X$ mediator to the SM as well as extra lighter particles present in the dark sector,  it can be much smaller than the typical WIMP case \cite{kindec}.  Thus, in our work, we restrict the kinetic decoupling temperature in a model-independent way by taking the limit from the Lyman-$\alpha$ forest, which excludes $T_{\rm kd}\lesssim 100\,{\rm eV}$. As a result, we obtain the upper bound on the relative velocity between $\phi_1$ and $\phi_2$ at recombination \cite{kai}, as follows,
\bea
v_{\rm rec} \lesssim 2\times 10^{-7} \bigg(1+\frac{m_2}{m_1}\bigg) \bigg(\frac{m_2}{100\,{\rm GeV}} \bigg)^{-1/2}.
\eea
Then, at such small velocities, the Planck data \cite{planck,planck15} sets the bound on the annihilation cross section in the following,
\bea
\langle\sigma v\rangle_{\phi_1\phi_2\to \phi_1 X}< 4\times 10^{-25}\,{\rm cm^3/s}\,\bigg(\frac{f_{\rm eff}}{0.1}\bigg)^{-1} \cdot\frac{1}{r_1(1-r_1)} \cdot\bigg(\frac{m_2}{100\,{\rm GeV}} \bigg) \label{CMB}
\eea
with $r_1=\Omega_1/\Omega_{\rm DM}$.
The parameter space with a strong Sommerfeld enhancement is constrained most by the CMB constraint \cite{kai}, so we discuss the benchmark models introduced in Figs.~\ref{relic} and \ref{relic2} in relation to the CMB constraint below.

 \begin{table}[hbt!]
  \begin{center}
  \scalebox{0.9}{
 \begin{tabular}{c|c|c|c|c|c|c|c|c|c}
      \hline\hline
      &&&&&&&&\\[-2mm]
      &   $m_2\simeq 2m_1$ &  $m_X$ & $\alpha$ & $\alpha_X$ & $\langle\sigma v\rangle^0_{\phi_1\phi_2\to \phi_1 X} $  & $r_1$ & $S_0$ & $\Delta$ & $\sigma_{\rm self}/m_{\rm eff}$ \\
           &   [GeV] &  [GeV] & $=\frac{g^2}{4\pi}$  & $=\frac{g^2_X}{4\pi}$  &  $[{\rm cm^3/s}]$  & $=\frac{\Omega_1}{\Omega_{\rm DM}}$  &  & $=1-\frac{m_2}{2m_1}$  & $[{\rm cm}^2/g]$ \\[2mm]
      \hline
      &&&&&&&&\\[-2mm]
      $\rm B1$ & 200 & 50& 0.05  & 0.0045 & $9.9\times 10^{-27}$  & $0.5$ & $444.7$  & $7.75\times10^{-4}$ & $0.014$
                     \\[2mm]
      \hline
      &&&&&&&&\\[-2mm]
 $\rm B2$ & 400  & 100  & 0.1  &  0.009 & $9.9\times 10^{-27}$   & $0.5$ & $889$  & $10^{-4}$ & $0.002$
                     \\[2mm]
                     \hline
      &&&&&&&&\\[-2mm]
 $\rm B3$ & 26 & 5  & 0.00032  & 0.04 &  $2.0\times 10^{-26}$  & $0.005$ & $1336$  & $5\times 10^{-10}$ & $0.003$
                     \\[2mm]
                     \hline
      &&&&&&&&\\[-2mm]
 $\rm B4$ & 240 &  60 & 0.0032  & 0.03 &  $2.9\times 10^{-27}$ & $0.075$ & $7379$  & $10^{-7.7}$ & $0.086$
                     \\[2mm]
      \hline\hline
    \end{tabular}}
  \end{center}
    \caption{Benchmark models and constraints on Sommerfeld factors}
      \label{table}
\end{table}

In Table~\ref{table}, we summarize four benchmark models, B1, B2, B3 and B4, taken in the previous subsection for  Figs.~\ref{relic} and \ref{relic2}. For each benchmark model, we list the semi-annihilation cross section for $\phi_1\phi_2\to \phi_1 X$ at tree level, the fraction of the dark matter abundance of $\phi_1$, $r_1$, the maximum value of the Sommerfeld factor  $S_0$ for $s$-wave,  the minimum value of the deviation, $\Delta=1-\frac{m_2}{2m_1}$, from the $u$-channel condition, and the effective self-scattering cross section,  $\sigma_{\rm self}/m_{\rm eff}$, from eq.~(\ref{scattrate}), for $\phi_1\phi_2\to \phi_1\phi_2$, in order.   All the benchmark models satisfy the CMB constraint given in eq.~(\ref{CMB}), and B4 model leads to a sizable self-scattering cross section with velocity dependence, that is marginally consistent to solve the small-scale problems \cite{smallscale3,diversity,sidm}.

\subsection{Direct detection for self-resonant dark matter} 

There are interesting signals from multi-component dark matter for direct detection experiments, depending on the mass differences between dark matter components \cite{boosted,BD,exodm0,exodm,ko}.
Since the $\phi_1$ dark matter particle in our model appearing from the semi-annihilation is boosted, it could lead to a new signature for direct detection experiments once $\phi_1$ is produced at the galactic center and it scatters strongly with the target material via nucleon or electron interactions.

In the galactic center of mass frame, the $\phi_1$ dark matter particle appears boosted due to the annihilation, $\phi_1\phi_2\to \phi_1 X$, with the same momentum as the one for the $X$ particle, as $p'_1=p_X$ in eq.~(\ref{Xmom}), and the energy given by
\bea
E'_1= \frac{m^2_2+2m_1 m_2+2m^2_1 -m^2_X}{2(m_1+m_2)},
\eea
leading to the gamma factor for the boosted dark matter $\phi_1$,
\bea
\gamma_1= \frac{E'_1}{m_1}= \frac{m^2_2+2m_1 m_2+2m^2_1 -m^2_X}{2m_1(m_1+m_2)}. \label{gamma1}
\eea
For instance, for the condition for the $u$-channel resonance, the above gamma factor becomes $\gamma_1=(10m^2_1-m^2_X)/(6m^2_1)$. 
Then, depending on the interactions between the $\phi_1$ particle and the SM, there is an interesting boosted signal in direct detection experiments such as XENON1T \cite{xenon1t,boosted,BD,exodm0,exodm,ko}. 

From the dark matter annihilation, $\phi_1\phi_2\to \phi_1 X$, the total flux for the boosted dark matter at the galactic center is given by
\bea
\Phi^{\rm G.C.}_1=1.6\times 10^{-4} {\rm cm^{-2}s^{-1}} \bigg( \frac{\langle\sigma v\rangle_{\phi_1\phi_2\to \phi_1 X}}{5\times 10^{-26} {\rm cm^3/s}}\bigg) \bigg(\frac{(1\,{\rm GeV})^2}{m_1 m_2}\bigg) r_1(1-r_1).  \label{BDflux}
\eea
Then, when the boosted dark matter scatters off the electron in the target material such as Xenon, the number of events is given \cite{BD} by
\bea
N_{\rm sig}= \frac{M T}{m_{\rm Xe}}\, \sigma_e\,\Phi^{\rm G.C.}_1 \label{electronsignal}
\eea
where $M$ is the target mass, $T$ is the exposure time,  $M/m_{X_e}$ is the number of Xenon atoms per ton, given by $4.2\times 10^{27}$, and $\sigma_e$ is the elastic scattering cross section between the boosted dark matter and electron. When we regard $X$ as being a heavy dark photon having a gauge kinetic mixing with the SM hypercharge gauge boson, the corresponding scattering cross section for electron is given \cite{exodm} by
\bea
\sigma_e&=&\frac{\varepsilon^2 e^2 g^2_X \mu^2_{e\phi_1}}{\pi m^4_X} \nonumber \\
&=& 10^{-33}\,{\rm cm}^2\bigg(\frac{\varepsilon}{10^{-3}}\bigg)^2\bigg( \frac{g_X}{0.1}\bigg)^2 \bigg(\frac{1\,{\rm GeV}}{m_X}\bigg)^4 \label{DMe}
\eea
where $\varepsilon$ is a dimensionless parameter determined by the gauge mixing parameter and $\mu_{e\phi_1}=m_e m_1/(m_e+m_1)$ is the reduced mass for the DM-electron system, which becomes $\mu_{e\phi_1}\simeq m_e$ for $m_{\phi_1}\gg m_e$.
For the XENON1T experiment with $MT=0.65\,{\rm ton-yr}$ \cite{xenon1t},  we need the DM-electron scattering cross section for the fixed number of events and the DM flux, as follows,
\bea
\sigma_e=10^{-33}\,{\rm cm}^2 \bigg(\frac{10^{-1}{\rm cm^{-2}s^{-1}} }{\Phi^{\rm G.C.}_1} \bigg) \bigg(\frac{N_{\rm sig}}{10}\bigg).
\eea
Then, in order to get $N_{\rm sig}\simeq 10$ from the flux for the boosted dark matter in eq.~(\ref{BDflux}), we would need the enhanced annihilation cross section to be $\langle\sigma v\rangle_{\phi_1\phi_2\to \phi_1 X}\simeq 10^{-22}\,{\rm cm^3/s}$ at the galactic center for $m_1=2m_2\sim 1\,{\rm GeV}$ and $r_1\simeq \frac{1}{2}$. In this case, the Sommerfeld factor from the $u$-channel resonance should be taken to $S_0\sim 10^4$. But, the electron/positron pair can be produced from the decay of the mediator $X$ for $m_X>2m_e$, so such a large cross section for $\langle\sigma v\rangle_{\phi_1\phi_2\to \phi_1 X}$ is constrained by gamma-ray  searches for light dark matter \cite{gammaray}. 

For the DM-electron scattering cross section, $\sigma_e\sim 10^{-33}\,{\rm cm}^2$,  the free stream length for the boosted dark matter inside the Earth is long enough, i.e., $L_{\rm fs}\sim 6000\,{\rm km}$,
for the averaged matter density, $\rho_E\sim 5.5\,{\rm g/cm^3}$ inside the Earth, so we can ignore the energy loss of the boosted dark matter inside the Earth.
The electron recoil energy due to the DM-electron scattering is bounded by $E_e=2m_e v^2_1$, so we can take $v_1=0.06$ for explaining the XENON1T electron excess  \cite{xenon1t} with the maximum recoil energy, $E^{\rm max}_e=3.6\,{\rm keV}$. In this case, from the boost factor  in eq.~(\ref{gamma1}) with $m_2=2m_1$, we need to take  $m_X=1.99729 m_1$.  Taking into account the DM-electron scattering cross section in eq.~(\ref{DMe}), dark matter particles and mediator masses should be of order GeV or below the GeV scale for the XENON1T electron excess. 

Instead of the boosted dark matter, we can also consider the direct detection for the $\phi_1$ particle present in the DM halo through the DM-electron scattering, but now with semi-conductors or superconducting detectors \cite{superdet}. 
But, the $u$-channel resonance is irrelevant for direct detection in this case, so we don't pursue this possibility further.

\section{Conclusions}

We made a comprehensive analysis of the new mechanism for Sommerfeld enhancement in the presence of $u$-channel resonances in models for multi-component dark matter with triple self-couplings. Assuming that dark matter is composed of at least two dark matter particles, we have shown that the elastic co-scattering for two dark matter particles can be enhanced by the $u$-channel pole from the lighter dark matter particle, when the two particle masses satisfy  $m_2\lesssim 2m_1$. 

We extended our previous analysis to higher partial waves for elastic co-scattering of dark matter and the effects of the $u$-channel resonances for $2\to 2$ and $3\to 2$ semi-annihilation process for multi-component dark matter, in the presence of a mediator particle.  In this case, the corresponding cross sections can be enhanced under the same mass relation, $m_2\lesssim 2m_1$, without a need of introducing a light mediator. 

In the effective theory for self-resonant dark matter, we also took various combinations of dark matter with spins and parities for elastic co-scattering and found that there are two-component dark matter models with scalar/scalar, scalar/pseudo-scalar, pseudo-scalar/fermion, vector/fermion dark matter that give rise to the $u$-channel Sommerfeld enhancements. 

We also considered the relic density condition for self-resonant dark matter in dark photon portal models and found several benchmark points where the abundances for two-component dark matter are the same or hierarchical, depending on  the annihilation strength of the lighter dark matter, in comparison to the one for the self-annihilation between two dark matter particles. In the case where the mediator particle in dark photon portal models couples to the SM, there are interesting indirect signals from the $2\to 2$ semi-annihilation with a large Sommerfeld enhancement in our model. Thus, we checked the consistency of the Sommerfeld factor for the $2\to 2$ semi-annihilation with indirect detection experiments such as CMB recombination and cosmic ray measurements. We also discussed the potential detection of a boosted dark matter in our model at direct detection experiments.

\section*{Acknowledgments}

HML thanks the organizers and participants in the Paris-Saclay Astroparticle Symposium 2021 for invitation and discussion.
The work is supported in part by Basic Science Research Program through the National Research Foundation of Korea (NRF) funded by the Ministry of Education, Science and Technology (NRF-2022R1A2C2003567 and NRF-2021R1A4A2001897). 
The work of SSK is supported by the Chung-Ang University Graduate Research Scholarship in 2020.

\end{document}